\pdfoutput=1
\documentclass[cernpreprint, atlasdraft=false, texlive=2023, orcidlogo, UKenglish, texmf, orcidlogo]{atlasdoc}
 
\usepackage[subfigure]{atlaspackage}
\usepackage{atlasbiblatex}
 
\usepackage{atlasphysics}
 
\graphicspath{{logos/}{figures/}}
 
\usepackage{ANA-TOPQ-2018-04-PAPER-defs}

\usepackage{multirow}

\AtlasTitle{Measurement of single top-quark production in the s-channel
in proton--proton collisions at $\mathrm{\sqrt{s}=13~TeV}$
with the ATLAS detector}
 
\AtlasAbstract{
A measurement of single top-quark production in the s-channel
is performed in proton--proton collisions at a centre-of-mass energy of $\SI{13}{TeV}$
with the ATLAS detector at the CERN Large Hadron Collider.
The dataset corresponds to an integrated luminosity of $\SI{139}{fb^{-1}}$.
The analysis is performed on events with an electron or muon,
missing transverse momentum and exactly two $b$-tagged jets in the final state.
A discriminant based on matrix element calculations is used
to separate single-top-quark s-channel events from the main background contributions,
which are top-quark pair production and $W$-boson production in association with jets.
The observed (expected) signal significance over the background-only hypothesis
is 3.3 (3.9) standard deviations, and the measured cross-section is
$\sigma=8.2^{+3.5}_{-2.9}\,\SI{}{pb}$,
consistent with the Standard Model prediction of
$\sigma^{\mathrm{SM}}=10.32^{+0.40}_{-0.36}\,\SI{}{pb}$.
}
 
\author{The ATLAS Collaboration}
 
\AtlasRefCode{TOPQ-2018-04}
\PreprintIdNumber{CERN-EP-2022-116}

\AtlasJournal{JHEP}
\AtlasJournalRef{JHEP 06 (2023) 191}
\AtlasDOI{10.1007/JHEP06(2023)191}

\hypersetup{pdftitle={ATLAS document},pdfauthor={The ATLAS Collaboration}}
 
\begin{document}
 
\maketitle
 
\tableofcontents

\section{Introduction}

\begin{figure}[!htb]
\centering
\includegraphics[width=0.32\textwidth]{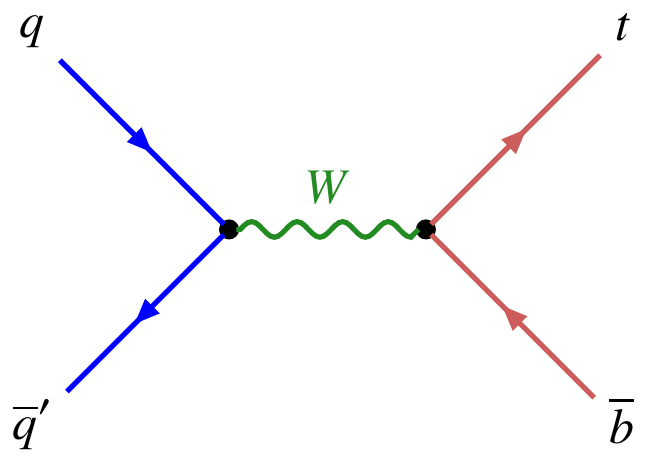}
\caption{
Feynman diagram for the dominant hard scattering process at leading-order in QCD of \schan single top-quark production.
The quarks in the initial (final) state are shown in blue (red),
and the exchanged \Wboson boson is shown in green.
}
\label{fig:feynman}
\end{figure}

In proton--proton (\pp) collisions, top quarks are produced predominantly in pairs via the strong interaction,
but also singly via the electroweak interaction through a $Wtb$ vertex.
Single top-quark production is therefore a powerful probe for the top quark electroweak couplings.
In the Standard Model (SM), three different production mechanisms are possible at leading-order (LO) in perturbative theory:
an exchange of a virtual \Wboson boson either in the \tchan or in the \schan,
or the associated production of a top quark and a \Wboson boson (\tW).
Figure~\ref{fig:feynman} shows the dominant Feynman diagram for \schan single top-quark production,
in which a top quark is produced with a bottom anti-quark in the final state.
This mode plays an important role in searches for new phenomena
that could be modelled as anomalous couplings or in effective field theories~\cite{Tait:2000sh,Cao:2007ea}.
 
Single top-quark production in proton--antiproton collisions was first observed by the CDF and D0 collaborations
in combined measurements of the \schan and \tchan~\cite{CDF:2009itk,D0:2009isq}.
The \schan production mode alone was later observed in a combination of the results from the two collaborations~\cite{CDF:2014uma}.
At the Large Hadron Collider (LHC), the production of single top quarks in \pp\ collisions was observed by the CMS and ATLAS collaborations
in \tchan production at a centre-of-mass energy of $\sqs=7$~\TeV~\cite{CMS-TOP-11-021,TOPQ-2011-14},
and in $tW$ associated production at $\sqs=8$~\TeV~\cite{CMS-TOP-12-040,TOPQ-2012-20}.
Both production modes were later measured at $\sqs=13$~\TeV~\cite{CMS-TOP-16-003,TOPQ-2015-15,CMS-TOP-20-002,TOPQ-2015-16}.
For \schan production, searches were performed by the CMS Collaboration, combining $\sqs=7$~\TeV\ and $\sqs=8$~\TeV\ data,
leading to an observed (expected) significance of 2.5 (1.1) standard deviations~\cite{CMS-TOP-13-009},
and by the ATLAS Collaboration with $\sqs=8$~\TeV\ data,
leading to an observed (expected) significance of 3.2 (3.9) standard deviations~\cite{TOPQ-2015-01},
over the background-only hypothesis.
 
In this paper, a measurement of the single-top-quark \schan production cross-section in \pp collisions at $\sqs=13$~\TeV\ is presented.
In the SM, the \schan single-top-quark cross-section in \pp collisions at $\sqs=13$~\TeV, calculated at next-to-leading order (NLO)
in quantum chromodynamics (QCD), is $\sigma^{\mathrm{SM}}=10.32^{+0.40}_{-0.36}\,\SI{}{pb}$ (cf. Section~\ref{sec:samples}).
The increase in the cross-section for this process between $\sqs=8$~\TeV\ and $\sqs=13$~\TeV\
is slightly larger than for \Wboson-boson production, which is the second dominant source of background;
it is however not as large as for \ttbar production, which is the dominant source.
Using predictions based on the same techniques as those described in Section~\ref{sec:samples},
the ratio of the \schan single-top-quark cross-section to the \Wboson-boson cross-section is $1.4\,10^{-4}$ ($1.7\,10^{-4}$),
and to the \ttbar cross-section is $2.1\,10^{-2}$ ($1.2\,10^{-2}$), at $\sqs=8$~\TeV\ ($\sqs=13$~\TeV).
This analysis is therefore more challenging as the centre-of-mass energy increases.
 
The top quark decays almost exclusively into a \Wboson boson and a $b$-quark.
This analysis considers only the leptonic decays (electron or muon) of the \Wboson boson,
because the fully hadronic final states are dominated by multijet background.
Some of the events containing a \Wboson boson decaying into a $\tau$-lepton which subsequently decays
leptonically are also selected.
At LO the final state contains two jets with large transverse momenta:
one jet containing $b$-hadrons (called a $b$-jet) originating from the decay of the top quark into a $b$-quark,
and another $b$-jet from the $Wtb$ vertex producing the top quark.
Thus the experimental signature consists of an isolated electron or muon,
large missing transverse momentum due to the undetected neutrino from the \Wboson-boson decay,
and two jets with large transverse momentum (\pT) which are both identified as containing $b$-hadrons ($b$-tagged).
To extract the signal events amidst the copious background, a `matrix element method'  discriminant~\cite{Kondo:1988yd,Kondo:1991dw} is used,
which assigns to each selected event a probability of being signal,
estimated by means of exact matrix element calculations at LO for signal and background processes.
This strategy has been shown to provide better sensitivity
than the boosted decision trees used in previous searches for s-channel production~\cite{TOPQ-2015-01,TOPQ-2014-03}.


\section{ATLAS detector}

\newcommand{\AtlasCoordFootnote}{
ATLAS uses a right-handed coordinate system with its origin at the nominal interaction point (IP)
in the centre of the detector and the \(z\)-axis along the beam pipe.
The \(x\)-axis points from the IP to the centre of the LHC ring,
and the \(y\)-axis points upwards.
Cylindrical coordinates \((r,\phi)\) are used in the transverse plane,
\(\phi\) being the azimuthal angle around the \(z\)-axis.
The rapidity is defined as $y=(1/2) \ln[(E+p_z)/(E-p_z)]$ where $E$
is the energy and $p_z$ is the longitudinal component of the momentum along the beam pipe.
The pseudorapidity is defined in terms of the polar angle \(\theta\) as \(\eta = -\ln \tan(\theta/2)\).
Angular distance is measured in units of \(\Delta R \equiv \sqrt{(\Delta\eta)^{2} + (\Delta\phi)^{2}}\).}
 
ATLAS~\cite{PERF-2007-01} is a multipurpose particle detector at the LHC
with a forward--backward symmetric cylindrical geometry and a near \(4\pi\) coverage in
solid angle.\footnote{\AtlasCoordFootnote}
It consists of an inner tracking detector surrounded by a thin superconducting solenoid
providing a \SI{2}{\tesla} axial magnetic field, electromagnetic and hadron calorimeters, and a muon spectrometer.
The inner tracking detector covers the pseudorapidity range \(|\eta| < 2.5\).
It consists of silicon pixel, silicon microstrip, and transition radiation tracking detectors.
Lead/liquid-argon (LAr) sampling calorimeters provide electromagnetic (EM) energy measurements
with high granularity.
A steel/scintillator-tile hadron calorimeter covers the central pseudorapidity range (\(|\eta| < 1.7\)).
The endcap and forward regions are instrumented with LAr calorimeters
for both the EM and hadronic energy measurements up to \(|\eta| = 4.9\).
The muon spectrometer surrounds the calorimeters and is based on
three large superconducting air-core toroidal magnets with eight coils each.
The field integral of the toroids ranges between \num{2.0} and \SI{6.0}{\tesla\metre}
across most of the detector.
The muon spectrometer includes a system of precision tracking chambers and fast detectors for triggering.
A two-level trigger system is used to select events.
The first-level trigger is implemented in hardware and uses a subset of the detector information
to accept events at a rate below \SI{100}{\kHz}.
This is followed by a software-based trigger that
reduces the accepted event rate to \SI{1}{\kHz} on average
depending on the data-taking conditions.
An extensive software suite~\cite{ATL-SOFT-PUB-2021-001} is used in the reconstruction and analysis of real
and simulated data, in detector operations, and in the trigger and data acquisition systems of the experiment.


\section{Data and simulation samples}
\label{sec:samples}


The data for this analysis were collected with the ATLAS detector at the LHC between 2015 and 2018,
in $pp$ collisions at $\sqrt{s}=13$~\TeV, using single-lepton triggers, and correspond to an integrated luminosity of 139~\ifb.
Only events for which the LHC beams were in stable-collision mode and all relevant subsystems were operational are
considered~\cite{DAPR-2018-01}.
 
Monte Carlo (MC) simulation is used to model signal and background processes.
Events were simulated using either the full ATLAS detector simulation~\cite{SOFT-2010-01}
based on \GEANT~\cite{Agostinelli:2002hh} or a faster simulation where the full \GEANT simulation
of the calorimeter response is replaced by a detailed parameterisation of the shower shapes~\cite{SOFT-2010-01}.
To simulate the effects of multiple interactions in the same and neighbouring bunch crossings (pile-up),
additional interactions were generated using \PYTHIA[8.186]~\cite{Sjostrand:2014zea}
with a set of tuned parameters called the A3 tune~\cite{ATL-PHYS-PUB-2016-017} and overlaid onto the simulated hard-scatter event.
Simulated events are reweighted to match the pile-up conditions observed in the full dataset,
and are processed through the same reconstruction algorithms and analysis chain as the data.
 
Events of single-top-quark \schan, \tchan, and \tW events, as well as \ttbar events,
were generated at NLO in QCD
using the \POWHEGBOX[v2] generator~\cite{Nason:2004rx,Frixione:2007vw,Alioli:2010xd,Alioli:2009je,Frederix:2012dh,Re:2010bp,Frixione:2007nw}
for the nominal prediction of these processes.
The top-quark mass was set to $\mtop=172.5\,\GeV$.
For single-top-quark s-channel and \tW production, and for \ttbar production, the samples were generated in the five-flavour scheme
and using the \NNPDF[3.0nlo] parton distribution function (PDF) sets~\cite{Ball:2014uwa},
and for single-top-quark \tchan production, the samples were generated in the four-flavour scheme and using the $\NNPDF[3.0nlo]\_$nf4 PDF sets.
The parton shower, hadronisation, and underlying event were added using \PYTHIA[8.230] with the
A14 tune~\cite{ATL-PHYS-PUB-2014-021} and the \NNPDF[2.3lo] PDF sets~\cite{Ball:2012cx}.
The decays of $b$- and $c$-hadrons were simulated using the \EVTGEN[1.6.0] program~\cite{Lange:2001uf}.
For \ttbar production, the \hdamp parameter\footnote{The \hdamp parameter is a resummation
damping factor and one of the parameters that control the matching of \POWHEG matrix elements
to the parton shower and thus effectively regulates the
high-\pt radiation against which the \ttbar system recoils.} in \POWHEGBOX was set to $1.5\cdot\mtop$.
For \tW production, the diagram removal (DR) scheme,
in which all doubly resonant NLO $tW$ diagram amplitudes are removed~\cite{Frixione:2008yi},
was employed to handle interference with \ttbar diagrams.
 
Alternative simulation samples are used to estimate modelling uncertainties of top-quark processes (cf.\ Section~\ref{sec:systematics}).
For the dominant \ttbar background process, two alternative samples are used:
one produced with \POWHEGBOX[v2] but with the \hdamp parameter set to $3\cdot\mtop$
in order to estimate the impact of varying the resummation, and the other produced with \MGNLO to assess
the effect of changing the matching of NLO matrix elements (MEs) to the parton shower.
For \tW production, another sample produced with \POWHEGBOX[v2] but using the diagram subtraction scheme,
where a gauge-invariant subtraction term modifies the NLO \tW cross-section to locally cancel out the \ttbar\ contribution~\cite{Frixione:2008yi},
is used to evaluate the impact of using a different algorithm to remove the overlap with \ttbar production.
For these three samples, \PYTHIA[8.230] is used for the parton shower and hadronisation.
Furthermore, for each of the four top-quark processes, another sample was produced with \POWHEGBOX[v2]
with the same settings as for the nominal \POWHEGBOX\!+\,\PYTHIA[8] samples, but with \HERWIG[7.04] instead of \PYTHIA[8.230] in order to assess the impact of using a different parton shower and hadronisation model.
These samples used the H7UE tune~\cite{Bahr:2008pv,Bellm:2015jjp}
and \MMHT[lo] PDF set \cite{Harland-Lang:2014zoa} with \HERWIG[7.04], and \EVTGEN[1.6.0] for $b$- and $c$-hadrons decays.
 
Samples of leptonic \Wboson- and \Zboson-boson (\Vboson) events where the boson decays leptonically,
and semileptonic diboson ($VV$) events where one boson decays leptonically and the other hadronically,
were simulated with the \SHERPA[2.2.1]~\cite{Bothmann:2019yzt} generator.
In the \Zboson-boson events samples, the invariant mass of the two leptons
are required to be larger than 40~\GeV.
For \Vboson production, MEs with NLO (LO) accuracy in QCD for up to two (four) parton emissions
were used,
while for \VV production, MEs with NLO (LO) accuracy for up to one (three) parton emission(s)
were used.
The MEs were calculated with the Comix~\cite{Gleisberg:2008fv}
and \OPENLOOPS~\cite{Buccioni:2019sur,Cascioli:2011va,Denner:2016kdg} libraries,
and matched to the \SHERPA parton shower~\cite{Schumann:2007mg} using the \MEPSatNLO
prescription~\cite{Hoeche:2011fd,Hoeche:2012yf,Catani:2001cc,Hoeche:2009rj}.
The \NNPDF[3.0nnlo] set of PDFs was used, along with a
dedicated set of tuned parton-shower parameters developed by the \SHERPA authors.
 
The simulated event samples above described are normalised to the integrated luminosity
corresponding to the analysed data sample, and using theory predictions for the
cross-sections of the individual processes, as described below.
The predicted cross-section for the \schan single-top-quark signal is
$\sigma^{\mathrm{SM}}=10.32^{+0.29}_{-0.24}\mathrm{(scales)}\pm0.27\mathrm{(PDF+\alphas)}\,\SI{}{pb}=10.32^{+0.40}_{-0.36}\mathrm{(total)}\,\SI{}{pb}$,
calculated at NLO in QCD with \HATHOR[2.1]~\cite{Aliev:2010zk,Kant:2014oha},
where the uncertainties due to PDFs and \alphas were calculated using the \PDFforLHC prescription~\cite{Butterworth:2015oua}
with the \MSTW[nlo] 68\% CL~\cite{Martin:2009iq,Martin:2009bu}, \CT[10nlo]~\cite{Lai:2010vv} and \NNPDF[2.3nlo]~\cite{Ball:2012cx} PDF sets,
and were added in quadrature to the effect of the scale uncertainty, defined by taking the envelope of the cross-section values
obtained by independently varying the renormalisation and factorisation scales upwards and downwards by a factor of 2
while preventing the two scales from differing by more than a factor of 2.
For \ttbar production, the predicted cross-section of $832^{+40}_{-46}\,\SI{}{pb}$ is used,
calculated at next-to-next-to-leading order (NNLO)
in QCD including the resummation of next-to-next-to-leading logarithmic (NNLL) soft-gluon terms with
\TOPpp[2.0]~\cite{Beneke:2011mq,Cacciari:2011hy,Baernreuther:2012ws,Czakon:2012zr,Czakon:2012pz,Czakon:2013goa,Czakon:2011xx},
where the uncertainty accounts for the effect of the PDF and \alphas uncertainties
calculated using the \PDFforLHC[15] prescription~\cite{Butterworth:2015oua}
with the \MSTW[nnlo]~\cite{Martin:2009iq,Martin:2009bu}, \CT[10nnlo]~\cite{Lai:2010vv,Gao:2013xoa}
and \NNPDF[2.3lo]~\cite{Ball:2012cx} PDF sets, summed in quadrature with the effect of the scale uncertainty.
The \tchan single-top-quark cross-section is $217^{+9}_{-8}\,\SI{}{pb}$,
calculated at NLO in QCD with \HATHOR[2.1]~\cite{Aliev:2010zk,Kant:2014oha},
where the uncertainty accounts for the effect of the PDF and \alphas uncertainties calculated
using the \PDFforLHC[15] prescription with the \MSTW[nlo] 68\% CL, \CT[10nlo],
and \NNPDF[2.3nlo] PDF sets, summed in quadrature with the effect of the scale uncertainty.
The \tW cross-section is calculated at NLO in QCD with NNLL soft-gluon corrections~\cite{Kidonakis:2010ux,Kidonakis:2013zqa},
giving a value of $71.7\pm3.8\,\SI{}{pb}$, where the uncertainty accounts for the effect
of the PDF uncertainty calculated using the \MSTW[nnlo] 90\% CL PDF set,
summed in quadrature with the effect of the scale uncertainty.
These predicted cross-sections for top-quark processes are calculated for a top-quark mass of $\mtop=172.5\,\GeV$.
The $V$+\,jets MC samples are normalised to cross-sections predicted at NNLO in QCD~\cite{Anastasiou:2003ds}:
$60.2\,\SI{}{nb}$ for $W(\to\ell\nu_{\ell})$\,+\,jets and $6.32\,\SI{}{nb}$ for $Z(\to\ell^+\ell^-)$\,+\,jets,
calculated using the \textsc{FEWZ} program~\cite{Melnikov:2006kv} with the \MSTW[nnlo] PDF set.
The diboson MC samples are normalised to the cross-section predicted by the \SHERPA[2.2.1]~\cite{Bothmann:2019yzt} generator.
 
To model the instrumental background from fake and non-prompt electrons (cf.\ Section~\ref{sec:backgrounds}),
dijet events were simulated using \PYTHIA[8.186] with the A14 tune
and the \NNPDF[2.3lo] PDF sets, and using the \EVTGEN[1.2.0] program to model the decays of bottom and
charm hadrons.
Here, $2\rightarrow2$ QCD processes were generated, including multijet,
$qg\rightarrow q\gamma$, $q\bar{q}  \rightarrow g\gamma$, electroweak (\Wboson/\Zboson) and \ttbar\ production processes.
This simulated sample was filtered at generator level to enrich the sample with jets
which are likely to resemble electron signatures in the detector:
events are kept if particles (excluding neutrinos and muons) deposit at least 17~\GeV\
of energy into a square area $\eta\times\phi=0.1\times0.1$,
mimicking the highly localised energy deposits characteristic of electrons.


\section{Event reconstruction and selection}
\label{sec:selection}


Events are required to have at least one primary vertex with two
or more tracks with $\pt > \SI{0.5}{\GeV}$. If more than
one vertex is found, the hard-scattering primary vertex is selected as the
one with the highest sum of squared transverse momenta of associated
tracks~\cite{PERF-2015-01}.
 
Events were recorded using single-lepton triggers with either a low \pt\
threshold and a lepton isolation requirement, or a higher threshold
but a looser identification criterion and without any isolation requirement.
The lowest \pt\ threshold in the single-muon trigger was 20 (26)~\GeV~\cite{TRIG-2018-01}
for data taken in 2015 (2016--2018), while in the single-electron trigger it was
24 (26)~\GeV~\cite{TRIG-2018-05}.
 
Electrons are reconstructed from tracks in the inner tracking detector (ID) associated with
topological clusters of energy depositions in the calorimeter~\cite{EGAM-2018-01}
and are required to have $\pt>\SI{10}{\GeV}$ and $|\eta|<2.47$.
Candidates in the calorimeter barrel--endcap transition region ($1.37 <|\eta| < 1.52$) are excluded.
Electrons must satisfy the \textit{Medium} likelihood identification criterion.
Muon candidates are identified by matching ID tracks to full tracks
or track segments reconstructed in the muon spectrometer,
using the \textit{Loose} identification criterion~\cite{MUON-2018-03}.
Muons are required to have $\pt>\SI{10}{\GeV}$ and $|\eta|<2.5$.
Lepton tracks must match the primary vertex of the event,
i.e.\ they have to satisfy $|z_0\sin(\theta)|<0.5$~mm and
$|d_0/\sigma(d_0)| < 5\,(3)$ for electrons (muons), where
$z_0$ is the longitudinal impact parameter relative to the primary vertex
and $d_0$ (with uncertainty $\sigma(d_0)$) is the transverse impact parameter
relative to the beam line.
 
Jets are reconstructed from noise-suppressed topological clusters of
calorimeter energy depositions~\cite{PERF-2014-07} calibrated at the
electromagnetic scale~\cite{JETM-2018-05}, using the \antikt~\cite{Cacciari:2008gp,Fastjet}
algorithm with a radius parameter of 0.4.
The average energy contribution from pile-up is subtracted
according to the jet area and jets are calibrated as described in
Ref.~\cite{JETM-2018-05} with a series of simulation-based corrections and in situ techniques.
Jets are required to satisfy $\pt > \SI{20}{\GeV}$ and $|\eta|<4.5$.
The effect of pile-up is reduced by a so-called jet vertex tagger (JVT) algorithm,
applied on jets with $\pt < \SI{120}{\GeV}$ and $|\eta|<2.5$, that uses tracking information
to reject calorimeter-based jets that are not consistent with originating from the primary vertex~\cite{PERF-2014-03}.
 
Jets containing $b$-hadrons are $b$-tagged in the $|\eta|<2.5$ range corresponding to the tracker acceptance,
with the MV2c10 multivariate algorithm~\cite{FTAG-2018-01},
which combines information about the transverse impact parameters of displaced tracks and the topological properties
of secondary and tertiary decay vertices reconstructed within the jet.
In this analysis, two working points are used,
defined by different thresholds for the MV2c10 discriminant,
and corresponding to 85\% and 77\% efficiencies
for $b$-jets with $\pt>\SI{20}{\GeV}$
as determined in simulated \ttbar events.
The corresponding rejection rate is 2.7 (4.9) for $c$-jets (containing $c$-hadrons and no $b$-hadrons),
and 25 (110) for light-jets (containing no $c$- or $b$-hadrons), for the 85\% (77\%) efficiency working point.
Correction factors are applied to the simulated events to compensate for differences between data and simulation in the
$b$-tagging efficiency for $b$-, $c$-, and light-jets~\cite{FTAG-2018-01,ATLAS-CONF-2018-001,ATLAS-CONF-2018-006}.
 
An overlap removal procedure is applied to prevent double-counting of objects.
The closest jet within $\Delta R_{y}=\sqrt{(\Delta y)^2+(\Delta \phi)^2}=0.2$
of a selected electron is removed.
If the nearest jet surviving that selection is within $\Delta R_{y} = 0.4$ of the electron, the electron is discarded.
Muons are usually removed if they are separated from the nearest jet by $\Delta R_{y} < 0.4$,
since this reduces the background from heavy-flavour decays inside jets.
However, if this jet has fewer than three associated tracks, the muon is kept and the jet is removed instead;
this avoids an inefficiency for high-energy muons undergoing significant energy loss in the calorimeter.
 
The missing transverse momentum (with magnitude \met) is reconstructed
as the negative vector sum of the \pt of all the selected electrons,
muons, and jets described above, with an extra `soft term'
built from additional tracks associated with the primary vertex, to make
it resilient to pile-up contamination~\cite{PERF-2016-07}.
 
Events are required to have exactly one lepton with $\pt > \SI{30}{\GeV}$,
and at least two jets with $\pt>\SI{25}{\GeV}$ and $|\eta|<2.5$.
The selected lepton is required to match a corresponding object at trigger level.
More stringent identification and isolation criteria are applied to increase background rejection:
events are kept if the selected electron (muon) satisfies the \textit{Tight} (\textit{Medium}) identification
and the \textit{Gradient}~\cite{EGAM-2018-01,PERF-2015-10} isolation criteria.
To reduce the contribution of multijet production, which is dominant at low transverse momentum
and low transverse mass of the \Wboson boson\footnote{The transverse mass of the \Wboson boson, \mtW, is computed from
the lepton transverse momentum, $\pT^{\ell}$ and the difference in azimuthal angle,
$\Delta\phi(\ell,\met)$ as $\mtW=\sqrt{2\met\pT^{\ell}[1-\cos(\Delta\phi(\ell,\met))]}$.} (denoted by \mtW),
events are required to have $\met>\SI{35}{\GeV}$ and $\mtW>\SI{30}{\GeV}$.
Events are categorised into four non-overlapping analysis regions, using additional selection criteria described in the following.
 
The signal region (\SR) targets the \schan single-top signal topology
and is used to perform the statistical analysis presented in Section~\ref{sec:signal_fit}.
In the \SR, events contain exactly two jets with $\pt>\SI{30}{\GeV}$ and $|\eta|<2.5$,
both $b$-tagged using the 77\% efficiency working point,
and at least one of them is required to have $\pt>\SI{40}{\GeV}$.
In order to reduce the contamination from the dilepton channel of \ttbar production (see Section~\ref{sec:backgrounds}),
events are rejected if they contain additional leptons with $\SI{10}{\GeV}<\pT<\SI{30}{\GeV}$,
regardless the more stringent identification and isolation criteria described in the previous paragraph.
To reduce the contamination from \tchan single-top production,
and from the semileptonic channel of \ttbar events,
events are rejected if they contain additional jets with
$\SI{20}{\GeV}<\pT<\SI{30}{\GeV}$ or with $2.5<|\eta|<4.5$.
 
The \Wjets validation region (\WVR) is enriched in \Wjets events by using a less stringent $b$-tagging requirement,
and is used to assess the modelling of the \Wjets background (cf.\ Section~\ref{sec:backgrounds}).
In the \WVR, events contain exactly two jets, both with $\pt>\SI{30}{\GeV}$ and $b$-tagged
using the 85\% efficiency working point, but at least one of these two jets must fail the 77\% working point requirement
in order to ensure that none of the selected events satisfies the \SR requirements.
Events are also required to contain no additional lepton with $\SI{10}{\GeV}<\pT<\SI{30}{\GeV}$
regardless the more stringent identification and isolation criteria.
 
Two validation regions are enriched in \ttbar events in order to assess the modelling of this process.
In the \ttVR[3] and \ttVR[4], events are required to have, respectively,
exactly three and exactly four jets with $\pt>\SI{25}{\GeV}$ and $|\eta|<2.5$, two of which must be
$b$-tagged using the 77\% efficiency working point.


\section{Background estimation}
\label{sec:backgrounds}

Several processes with a final state that is identical or very similar to the single-top \schan
signal are sources of background events for this analysis.
The largest background source is \ttbar production, which is difficult to distinguish
from the signal since such events contain real top-quark decays.
In the dileptonic decay mode, \ttbar events can mimic the final-state signature of the signal
if one of the two leptons is unidentified, whereas the semileptonic decay mode contributes
to the selected samples if only two jets are reconstructed.
The second most important background source is \Wjets production.
Such events have the same signature as the signal if two jets produced
in association with the \Wboson boson are $b$-jets.
Furthermore, \Wboson-boson production in association with $c$- or light-jets
contributes due to misidentification of such jets as $b$-jets;
this represents 3\% of the expected \Wjets yield.
Single top-quark production in the \tchan also leads to a sizeable background contribution,
while associated \tW production has a smaller event yield.
Both \Zjets and diboson production are minor background sources.
These background processes, as well as the signal, are modelled using MC event samples and normalised to cross-sections
predicted in the SM, as described in Section~\ref{sec:samples},
before the signal extraction fit described in Section~\ref{sec:signal_fit}.

In addition, multijet production contributes to the selected events, when jets, non-prompt leptons from heavy-flavour decays,
or electrons from photon conversion, are misidentified as prompt isolated leptons.
The modelling of this background in the electron (muon) channel relies on the so-called jet-electron (anti-muon)
method described in Refs.~\cite{ATLAS-CONF-2014-058,TOPQ-2015-05}, in which a dedicated simulated (data) event sample is used as a
model for the shapes of the different distributions.
The normalisation of the multijet background is extracted from data for both channels,
using a dedicated fit described in this section.
Both normalisations and shapes are later used as input for the signal
extraction fit described in Section~\ref{sec:signal_fit}.
 
In the case of the jet-electron model, a dedicated selection is imposed
on the MC simulated dijet events sample described in Section~\ref{sec:samples},
in order to select the events with jets that are likely to resemble a lepton in the detector.
Such jets are required to deposit a large amount of their energy in the electromagnetic calorimeter,
and are treated as electrons in the event selection.
In order to reduce the uncertainty due to the size of the MC sample,
the $b$-tagging requirements imposed on two of the jets
in the event selection described in Section~\ref{sec:selection} are not applied.
In the case of the anti-muon model, a dedicated data sample is used.
Some of the muon identification criteria, related to the energy loss in the calorimeter,
to the longitudinal impact parameter, or to the tracking and calorimeter isolation,
are inverted or changed~\cite{ATLAS-CONF-2014-058,MUON-2018-03},
resulting in a sample that is enriched with events containing fake or non-prompt muons from multijet events.
 
The jet-electron and anti-muon event samples are then used to estimate the multijet background normalisation
in a binned maximum-likelihood fit, using variables chosen to provide good separation
between the background sources: the \MET distribution in the electron channel,
and the \mtW distribution in the muon channel.
For this purpose, the region definitions are modified
in order to increase the number of multijet background events in the selected sample:
the requirements on \MET in the electron channel, and on \mtW in the muon channel, are not applied.
A simultaneous fit of the electron and muon channels is performed in the \SR
to extract the normalisations of both the jet-electron and anti-muon samples,
which are allowed to float freely and independently in the fit.
The obtained multijet normalisation is used in the \SR as an initial value for the signal extraction fit
described in Section~\ref{sec:signal_fit}.
The normalisation of the \Wjets contribution is also allowed to float freely in the fit,
and so is the normalisation of the top-quark processes.
The latter are grouped together in one template where the relative contributions are fixed
to their SM predictions, as they show similar \MET and \mtW distributions.
The normalisation factors extracted for the \Wjets and top-quark processes in this fit
are, however, not used for the signal extraction fit described in Section~\ref{sec:signal_fit}.
The normalisations of the other processes are fixed to their SM predictions.
The fit is repeated in each of the three VRs to extract the multijet normalisation in these regions.
In the \WVR (\ttVR[3] and \ttVR[4]), the normalisation of the \Wjets contribution (top-quark processes contribution)
is allowed to float freely, while the normalisations of the other processes are fixed to their SM predictions.
The multijet normalisation obtained in the three VRs is used
to validate the modelling of the discriminant, as described in Section~\ref{sec:mem}.
 
Figures~\ref{fig:QCDfit_el} and~\ref{fig:QCDfit_mu} show the \MET and \mtW distributions in the electron and muon
channels respectively, after the fit of the multijet background in each region.
In these distributions, the uncertainties in the background normalisations
described in Section~\ref{sec:systematics} are included,
as well as \SI{6}{\%} and \SI{40}{\%} uncertainties in the \ttbar and \Wjets background normalisations,
motivated respectively by the uncertainty in the \ttbar cross-section prediction (cf.\ Section~\ref{sec:samples})
and the value of the \Wjets normalisation factor extracted in the fit --
while they are eventually treated as unconstrained parameters 
in the statistical analysis of the signal cross-section (cf.\ Section~\ref{sec:signal_fit}).
The predictions after the fit are in agreement with the data, given the statistical and systematic uncertainties.
The trends seen at low \MET or \mtW values for the best fit distribution would not impact the signal extraction fit,
for which the $\met>\SI{35}{\GeV}$ and $\mtW>\SI{30}{\GeV}$ requirements are applied (cf.\ Section~\ref{sec:selection}).
Furthermore, the trends seen at high \MET or \mtW values are consistent with the mismodelling, seen in other analyses~\cite{TOPQ-2020-02},
of \ttbar\ events by MC generators at NLO in QCD, which predict harder top-quark \pt spectra than observed in data,
and which are typically covered by the relevant \ttbar\ modelling systematic uncertainties described in Section~\ref{sec:systematics}.

\begin{figure}[htbp]
\centering
\subfigure[]{
\includegraphics[width=0.4\textwidth]{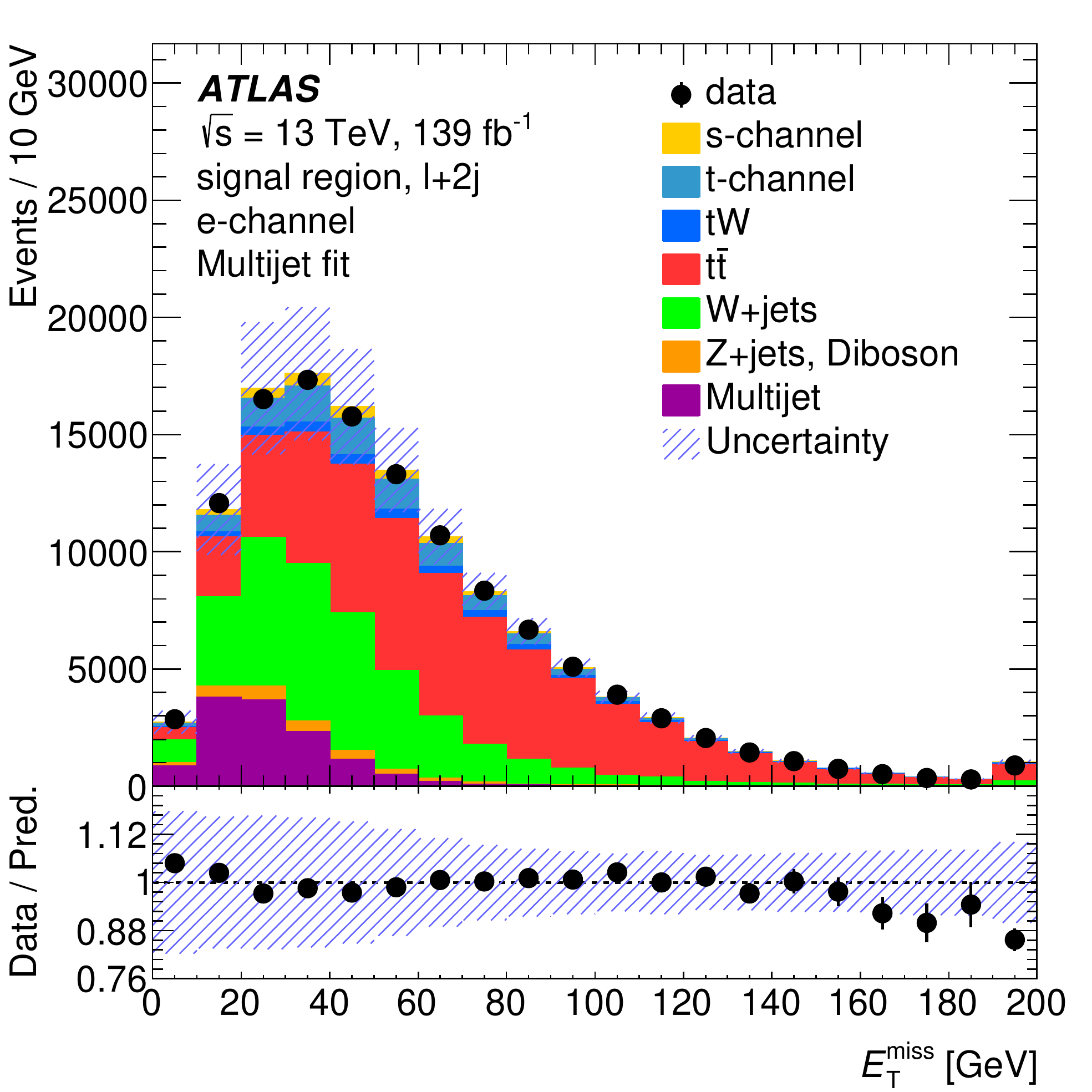}
\label{subfig:SR-el-met}
}
\subfigure[]{
\includegraphics[width=0.4\textwidth]{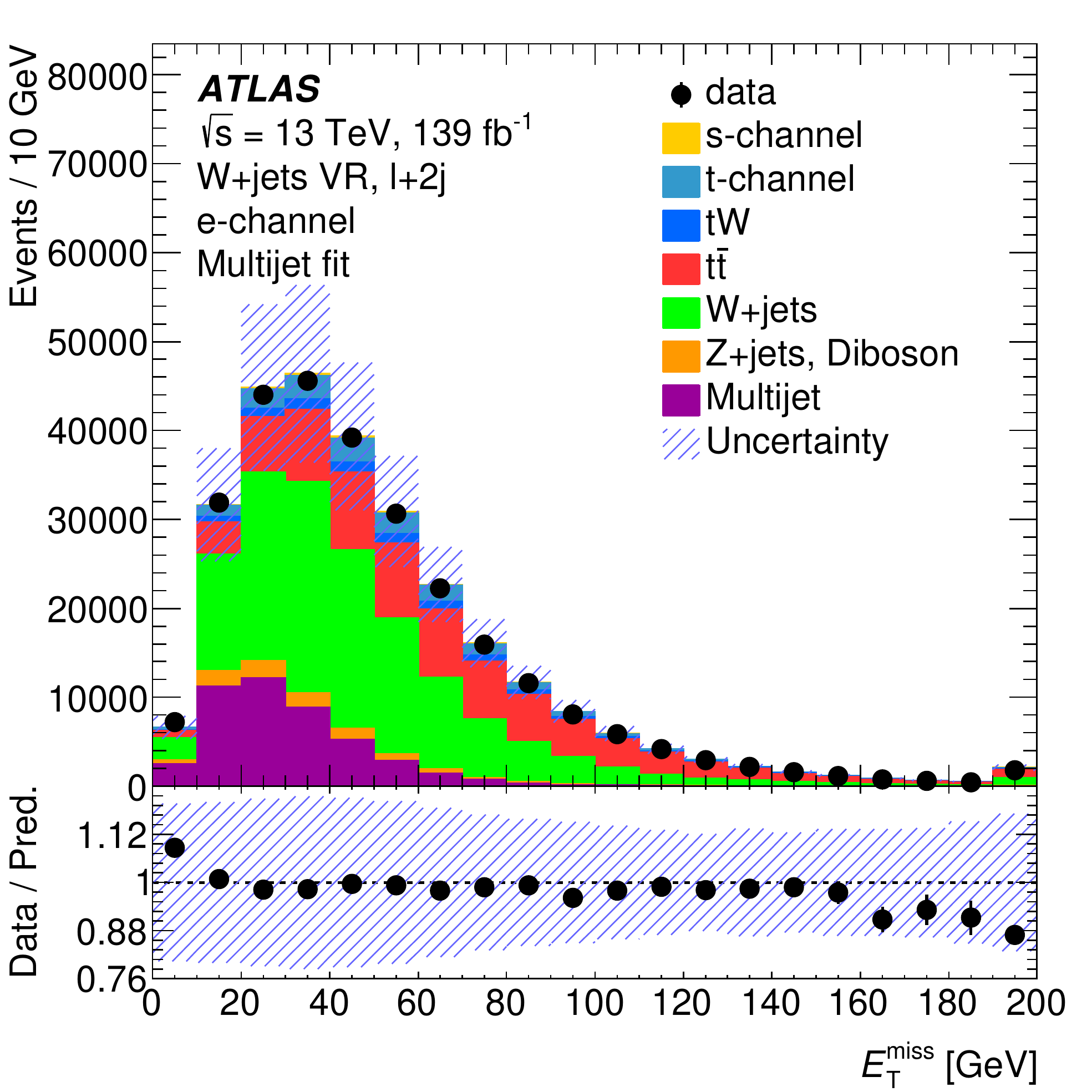}
\label{subfig:wjetsVR-el-met}
}\\
\subfigure[]{
\includegraphics[width=0.4\textwidth]{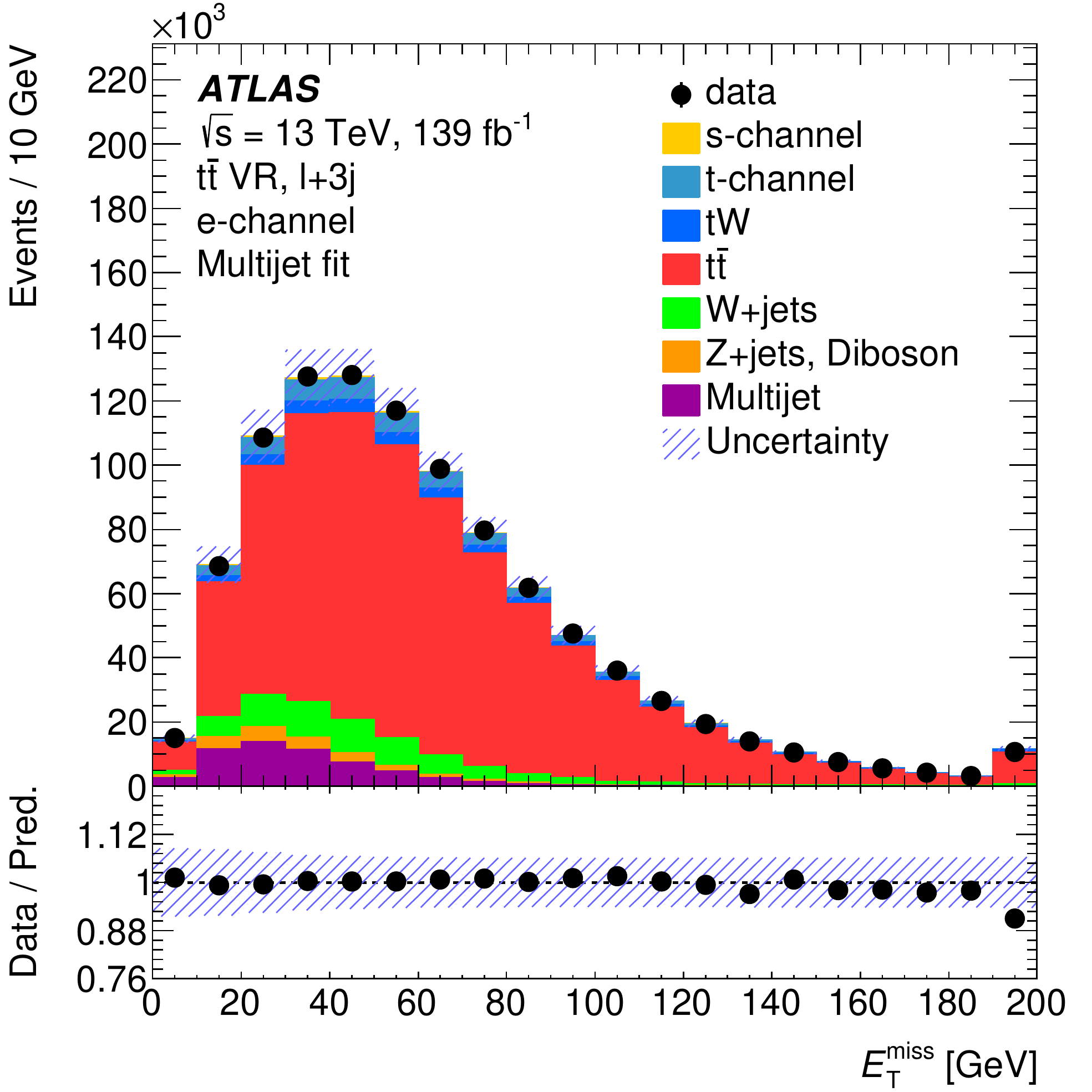}
\label{subfig:ttbar3jetVR-el-met}
}
\subfigure[]{
\includegraphics[width=0.4\textwidth]{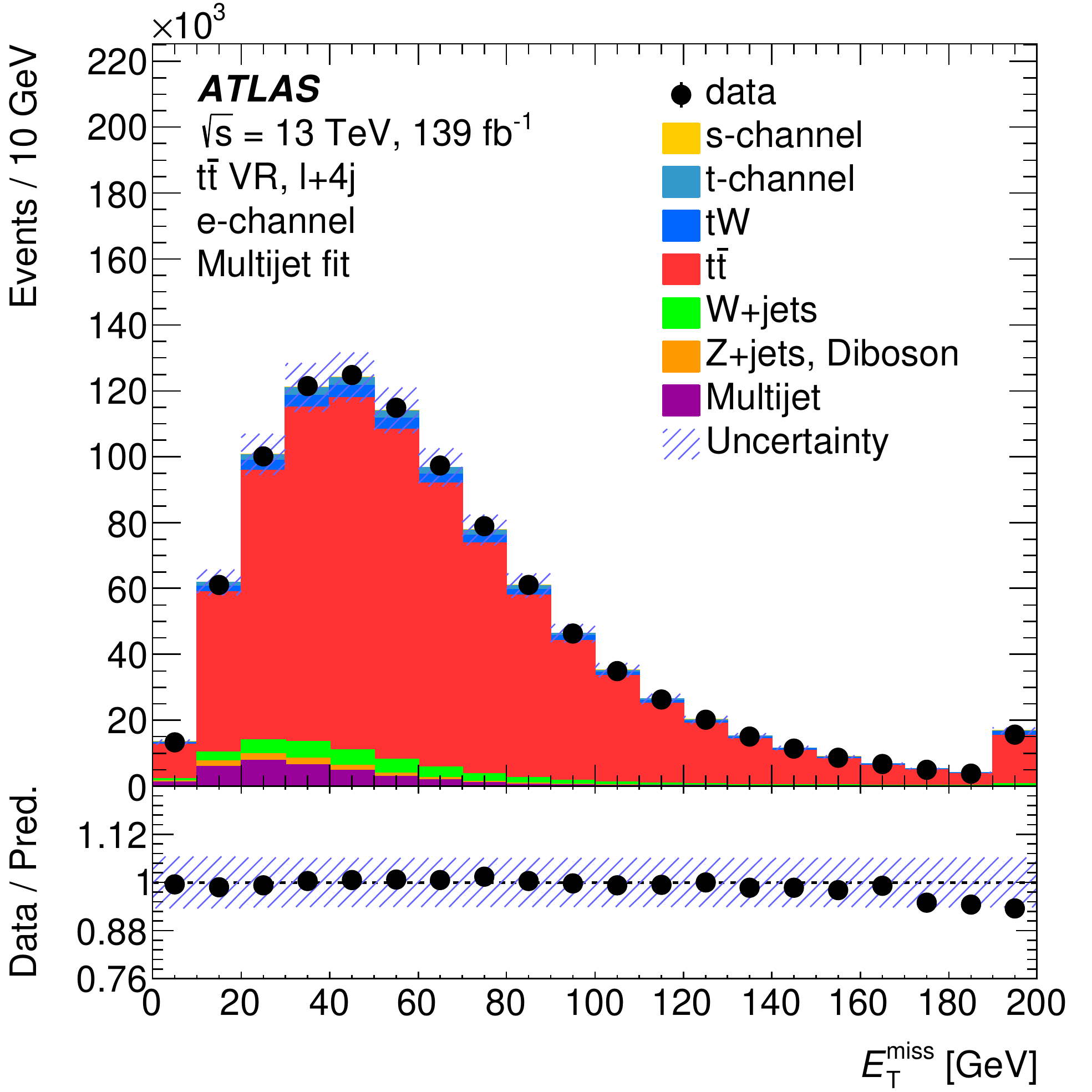}
\label{subfig:ttbar4jetVR-el-met}
}
\caption{
Distribution of \MET after the fit of the multijet backgrounds, in the electron channel,
in the \subref{subfig:SR-el-met}~signal region, \subref{subfig:wjetsVR-el-met}~\WVR,
\subref{subfig:ttbar3jetVR-el-met}~\ttVR[3], and \subref{subfig:ttbar4jetVR-el-met}~\ttVR[4] without applying the requirement on \MET.
Simulated events are normalised to the expected number of events given the integrated luminosity, after applying the normalisation factors obtained in the multijet fit.
The last bin includes the overflow.
The uncertainty bands indicate the simulation's statistical uncertainty,
the normalisation uncertainties for different processes ($40\%$ for \Wjets production, $30\%$ for multijet background and $6\%$ for top-quark processes)
and the multijet background shape uncertainty in each bin, summed in quadrature.
The lower panels show the ratio of the data to the prediction.
}
\label{fig:QCDfit_el}
\end{figure}
 
\begin{figure}[htbp]
\centering
\subfigure[]{
\includegraphics[width=0.4\textwidth]{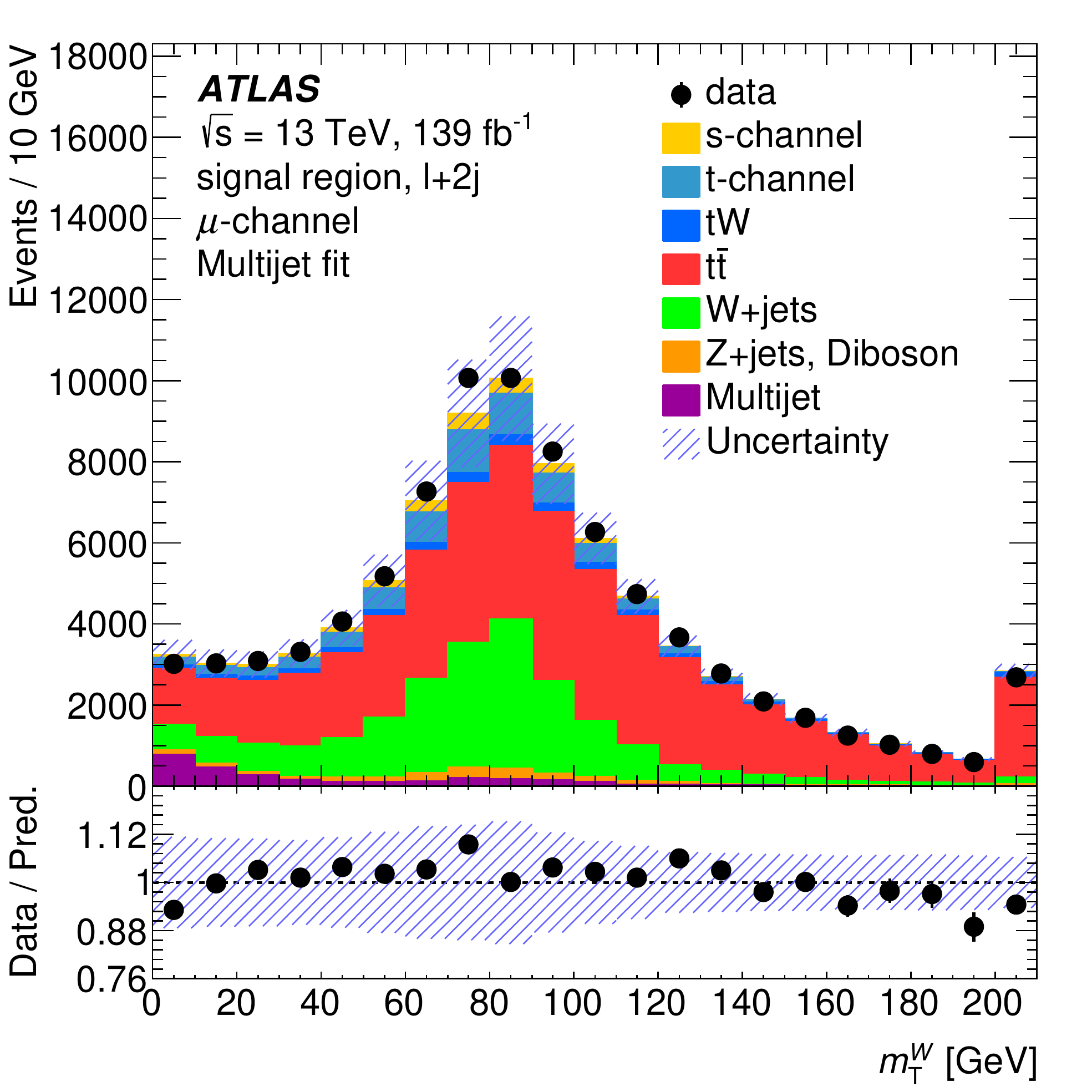}
\label{subfig:SR-el-mtw}
}
\subfigure[]{
\includegraphics[width=0.4\textwidth]{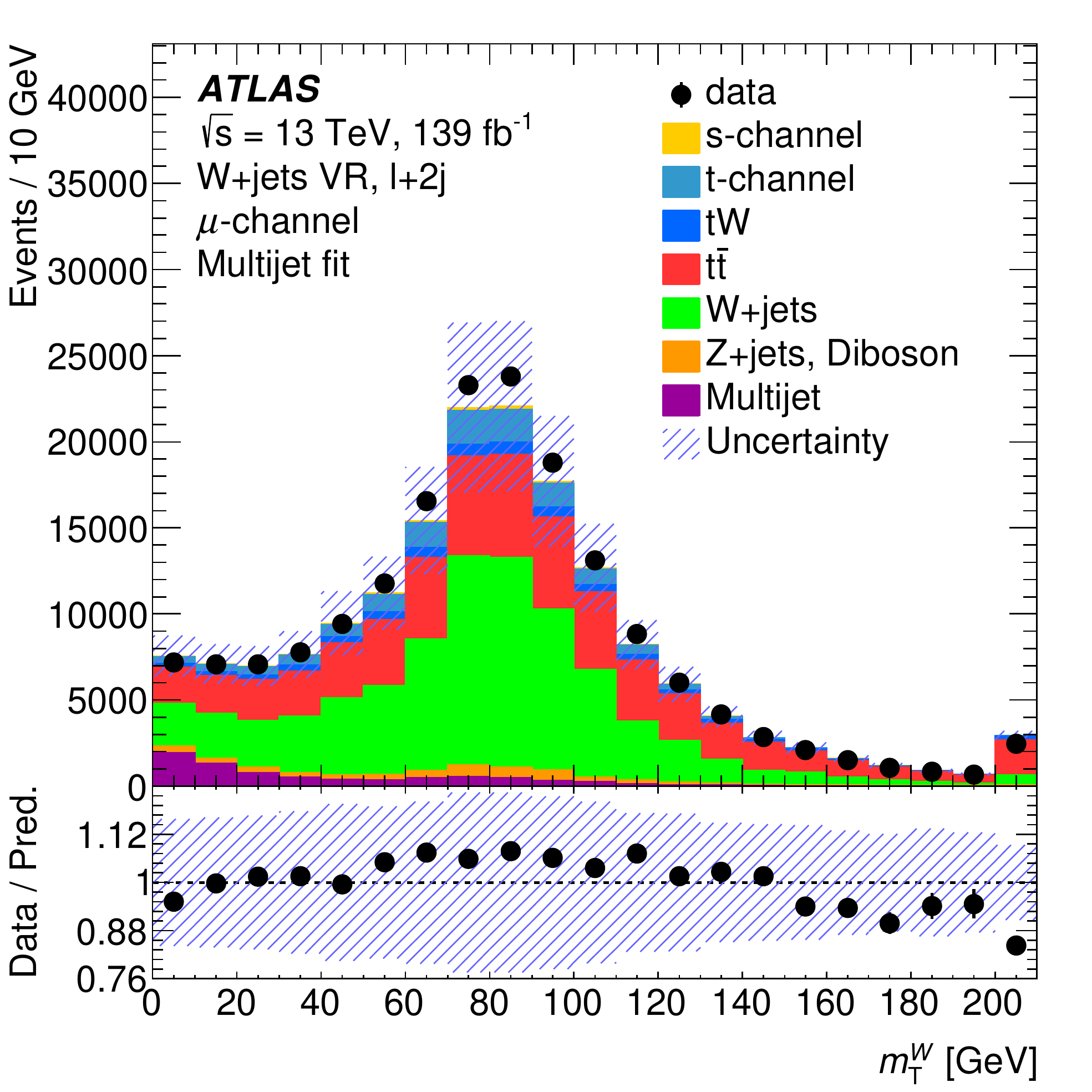}
\label{subfig:wjetsVR-el-mtw}
}\\
\subfigure[]{
\includegraphics[width=0.4\textwidth]{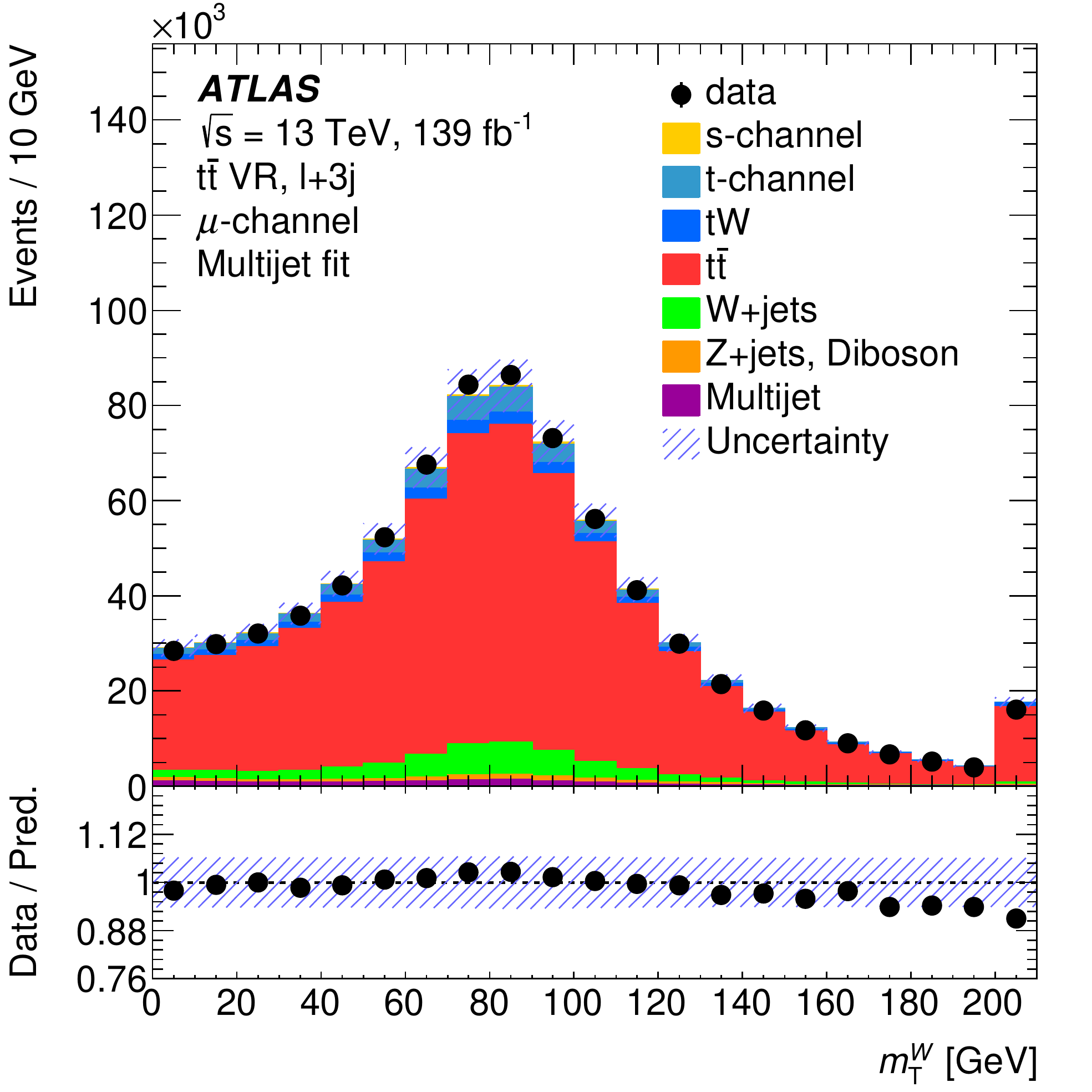}
\label{subfig:ttbar3jetVR-el-mtw}
}
\subfigure[]{
\includegraphics[width=0.4\textwidth]{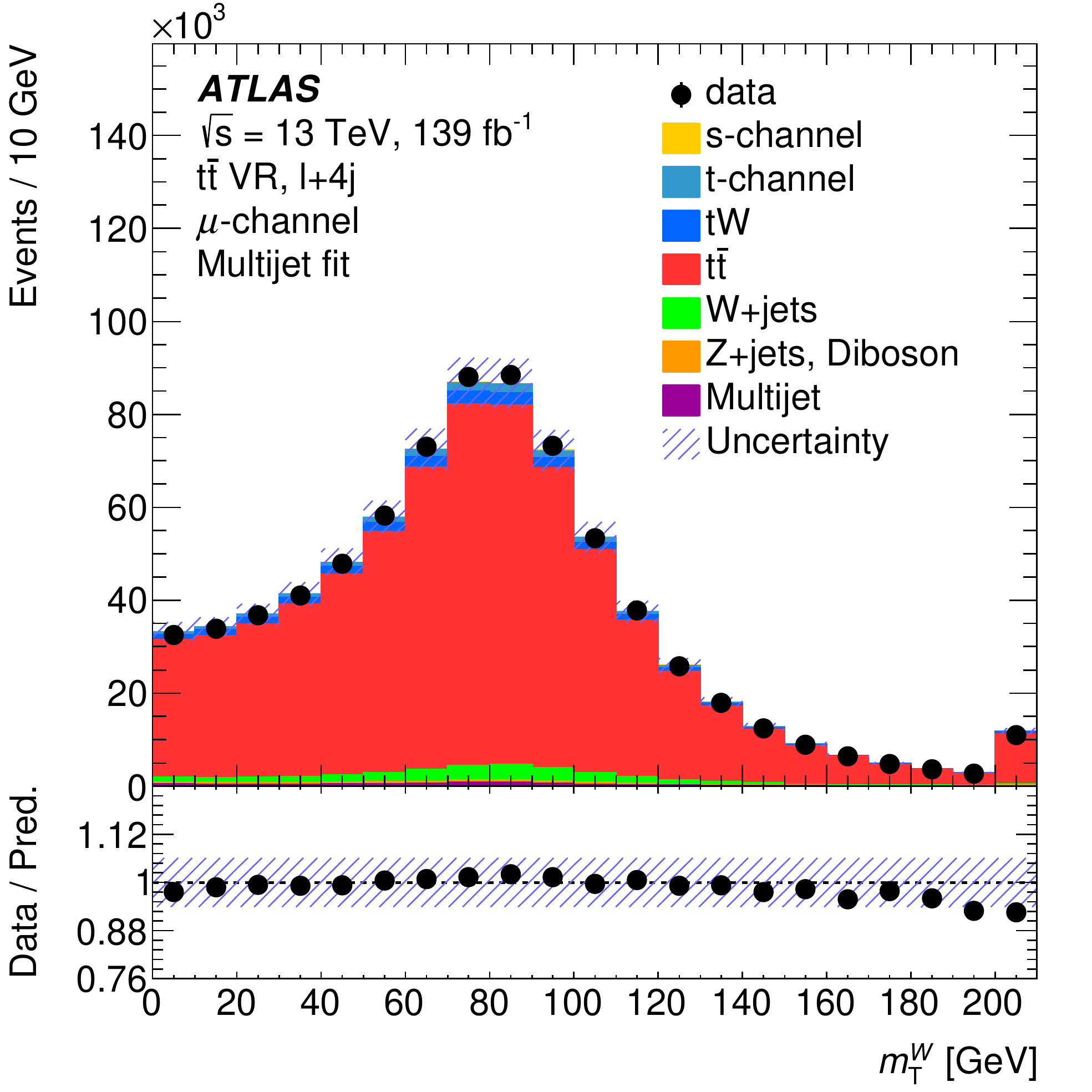}
\label{subfig:ttbar4jetVR-el-mtw}
}
\caption{
Distribution of \MTW after the fit of the multijet backgrounds, in the muon channel,
in the  \subref{subfig:SR-el-mtw}~signal region, \subref{subfig:wjetsVR-el-mtw}~\WVR,
\subref{subfig:ttbar3jetVR-el-mtw}~\ttVR[3], and \subref{subfig:ttbar4jetVR-el-mtw}~\ttVR[4] without applying the requirement on \MTW.
Simulated events are normalised to the expected number of events given the integrated luminosity, after applying the normalisation factors obtained in the multijet fit.
The last bin includes the overflow.
The uncertainty bands indicate the simulation's statistical uncertainty,
the normalisation uncertainties for different processes ($40\%$ for \Wjets production, $30\%$ for multijet background and $6\%$ for top-quark processes)
and the multijet background shape uncertainty in each bin, summed in quadrature.
The lower panels show the ratio of the data to the prediction.
}
\label{fig:QCDfit_mu}
\end{figure}


\section{Matrix element method}
\label{sec:mem}

The matrix element method (MEM)~\cite{Kondo:1988yd,Kondo:1991dw} directly uses theoretical calculations to compute a per-event signal probability.
This technique was used for the observation of single top-quark production in proton--antiproton collisions
at the Tevatron~\cite{CDF:2009itk,D0:2009isq,CDF:2008yej,D0:2006ngk,D0:2008wma}, and in finding evidence of single top-quark \schan production
in proton--proton collisions in ATLAS~\cite{TOPQ-2015-01}.
The discrimination between signal and background is based on the computation of likelihood values $\mathcal{P}(X\,|\ H_{\mathrm{proc}})$
for the hypothesis that a measured event with final state $X$
defined by the presence of certain reconstructed objects, is of a certain process type $H_{\mathrm{proc}}$.
The likelihoods can be computed by means of the factorisation theorem from the corresponding partonic
cross-sections of the hard scattering process, as follows:
\begin{align*}
\mathcal{P}(X\,|\ H_{\mathrm{proc}})  = &\ \int\mathrm{d}\Phi\,\frac{1}{\sigma_{H_{\mathrm{proc}}}}\frac{\mathrm{d}\sigma_{H_{\mathrm{proc}}}}{\mathrm{d}\Phi}T_{H_{\mathrm{proc}}}(X\,|\ \Phi)\ .
\end{align*}
The normalised fully differential cross-section $(1/\sigma_{H_{\mathrm{proc}}})(\mathrm{d}\sigma_{H_{\mathrm{proc}}}/\mathrm{d}\Phi)$
gives the probability density for a scattering process $H_{\mathrm{proc}}$ to lead to a parton-level final state $\Phi$
as a function of the four-momenta of all outgoing particles.
The mapping between the measured final state $X$ and the parton-level state $\Phi$
is implemented by transfer functions $T_{H_{\mathrm{proc}}}(X\,|\ \Phi)$ which take into account the detector energy-resolution functions,
the reconstruction efficiencies for the electron, the muon and the jets, and the $b$-tagging efficiencies,
as a function of the parton-level transverse momenta and pseudorapidities,
and the efficiency of the \MET selection as a function of the neutrino transverse momentum.
The permutations between the partons and the reconstructed objects are taken into account
by summing over the possible configurations which contribute to the differential cross-section~\cite{CDF:2008yej,D0:2008wma}.
 
The phase-space integration of the differential partonic cross-sections is performed by using the Monte Carlo
integration algorithm Vegas~\cite{Lepage:1977sw} from the Cuba program library~\cite{Hahn:2005pf}.
The required PDF sets are taken from the LHAPDF5 package~\cite{Butterworth:2014efa},
while the computation of the scattering amplitudes is based on code
from the MCFM program~\cite{Campbell:2010ff}.
The functional forms and parameterisations of the ATLAS detector resolutions used for the
transfer functions are those used in the KLFitter kinematic fit framework~\cite{TOPQ-2011-08,Erdmann:2013rxa}.
The parameters are derived from event samples produced using the ATLAS detector simulation.
 
In total, eight different processes, all having at least one charged lepton, one neutrino, and two quarks in the final state,
are considered for the computation of the likelihood values:
two processes for the \schan single-top-quark signal (one with two final-state partons corresponding to the
$b$-quarks in the LO diagram, and one with three final-state partons in which a real radiation correction is included),
one process for the \tchan single-top-quark background (which is modelled in the four-flavour scheme only),
two processes for \ttbar production (for the semileptonic and dileptonic final states, which are evaluated separately),
and three processes for the remaining \Wboson-boson background production (one process with two associated light-jets,
one process with one light-jet and one $c$-jet, and one process with two $b$-jets).
The scattering matrix elements for all considered processes are calculated using LO amplitudes.
 
From the likelihood values of these processes the probability $P(S\,|X)$ for a measured event $X$ to be a signal
event $S$ can be computed with Bayes' theorem by:
\begin{equation*}
P(S\,|\ X) = \frac{\sum_i P(S_i)\,\mathcal{P}(X\,|\ S_i)}{\sum_i P(S_i)\,\mathcal{P}(X\,|\ S_i)+\sum_j P(B_j)\,
\mathcal{P}(X\,|\ B_j)}\ .
\end{equation*}
Here, $S_i$ and $B_j$ denote all signal and background processes that are considered.
The a priori probabilities $P(S_i)$ and $P(B_j)$ are
taken to be the expected event fractions from the various processes contributing to the signal region, as obtained from Monte Carlo simulations;
variations of these factors were observed to have negligible impact on the sensitivity.
The final state $X$ of the measured events have the \SR topology: one electron or muon, two $b$-tagged jets, and missing transverse momentum.
The value of $P(S\,|X)$ is taken as the discriminant in the signal extraction.
Figure~\ref{fig:mem-separation} shows its expected distribution for the \schan single-top-quark signal
and for the \ttbar\ and \Wjets\ background contributions in the SR,
with the same binning and range as used for the signal extraction described in Section~\ref{sec:signal_fit}.
 
\begin{figure}[!htb]
\centering
\includegraphics[width=0.4\textwidth]{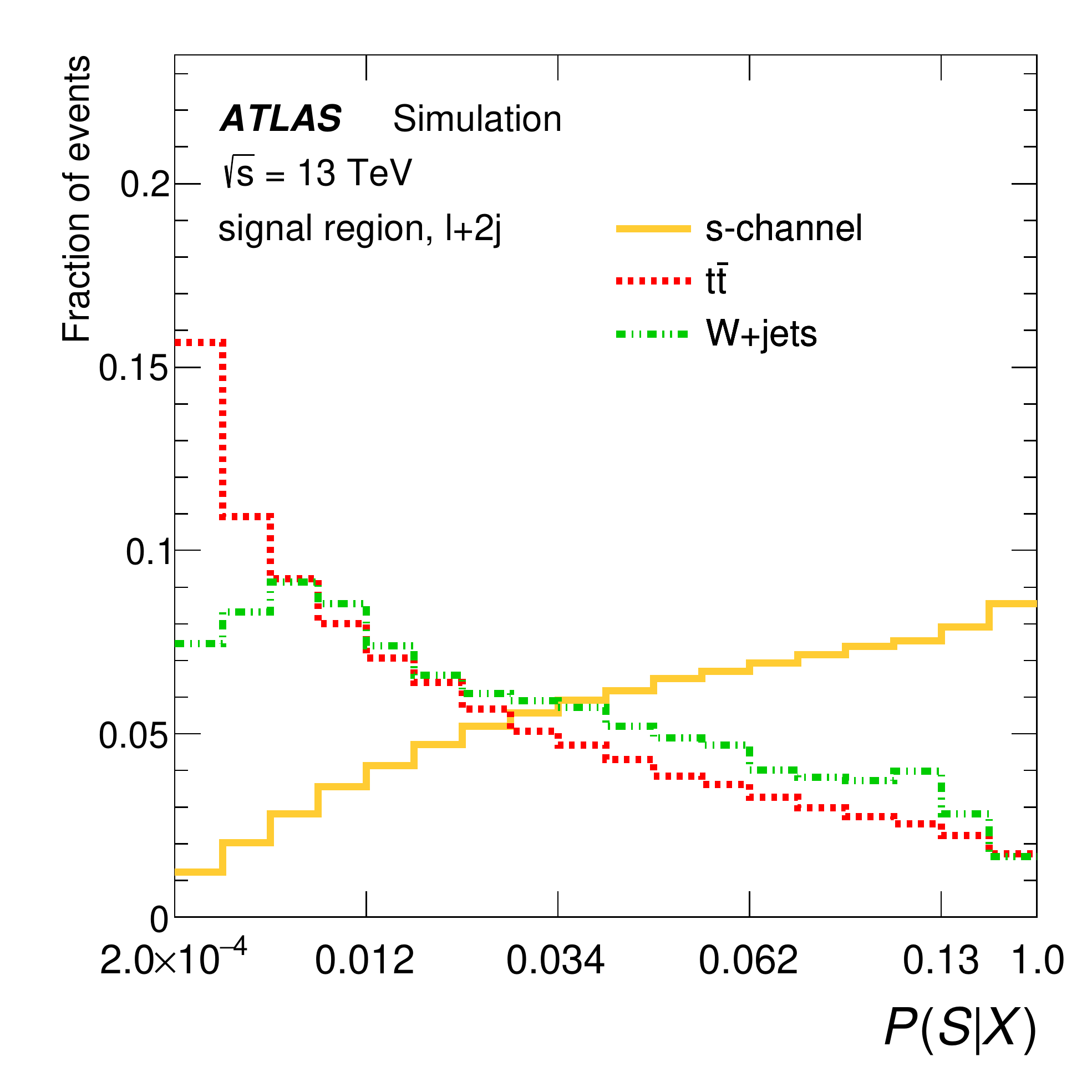}
\caption{
Expected distributions of the MEM discriminant $P(S|X)$ in the SR, for the \schan single-top signal,
and for the \ttbar\ and \Wjets\ backgrounds, for MEM discriminant values larger than $2.0\times10^{-4}$.
Each distribution is normalised to unity.
The binning is the same as the optimised binning used in the signal extraction fit, resulting in a non-linear horizontal scale.
}
\label{fig:mem-separation}
\end{figure}
 
In order to validate the \schan MEM discriminant $P(S\,|X)$, it is computed in the three validation regions.
For the \ttbar\ VRs, in which events with three or four jets are selected, only the two $b$-tagged jets
are considered for the MEM discriminant computation,
since most of the considered processes have only two outgoing partons.
For the jet-electron sample, in which the $b$-tagging requirement is not applied (cf.\ Section~\ref{sec:backgrounds}),
only the two leading jets are considered.
Figure~\ref{fig:mem-disc-validation} compares the discriminant distribution in data with that in simulation
for the three VRs, using the same optimised binning as used
for the signal extraction fit in the \SR described in Section~\ref{sec:signal_fit}.
In these distributions, produced prior to the signal extraction fit, the uncertainty band includes the normalisation uncertainties of all processes, including the signal and the \ttbar\ and \Wjets\ backgrounds, as well as the simulation's statistical uncertainty.
Good modelling is observed in the three validation regions.
 
\begin{figure}[!htb]
\centering
\subfigure[]{
\includegraphics[width=0.4\textwidth]{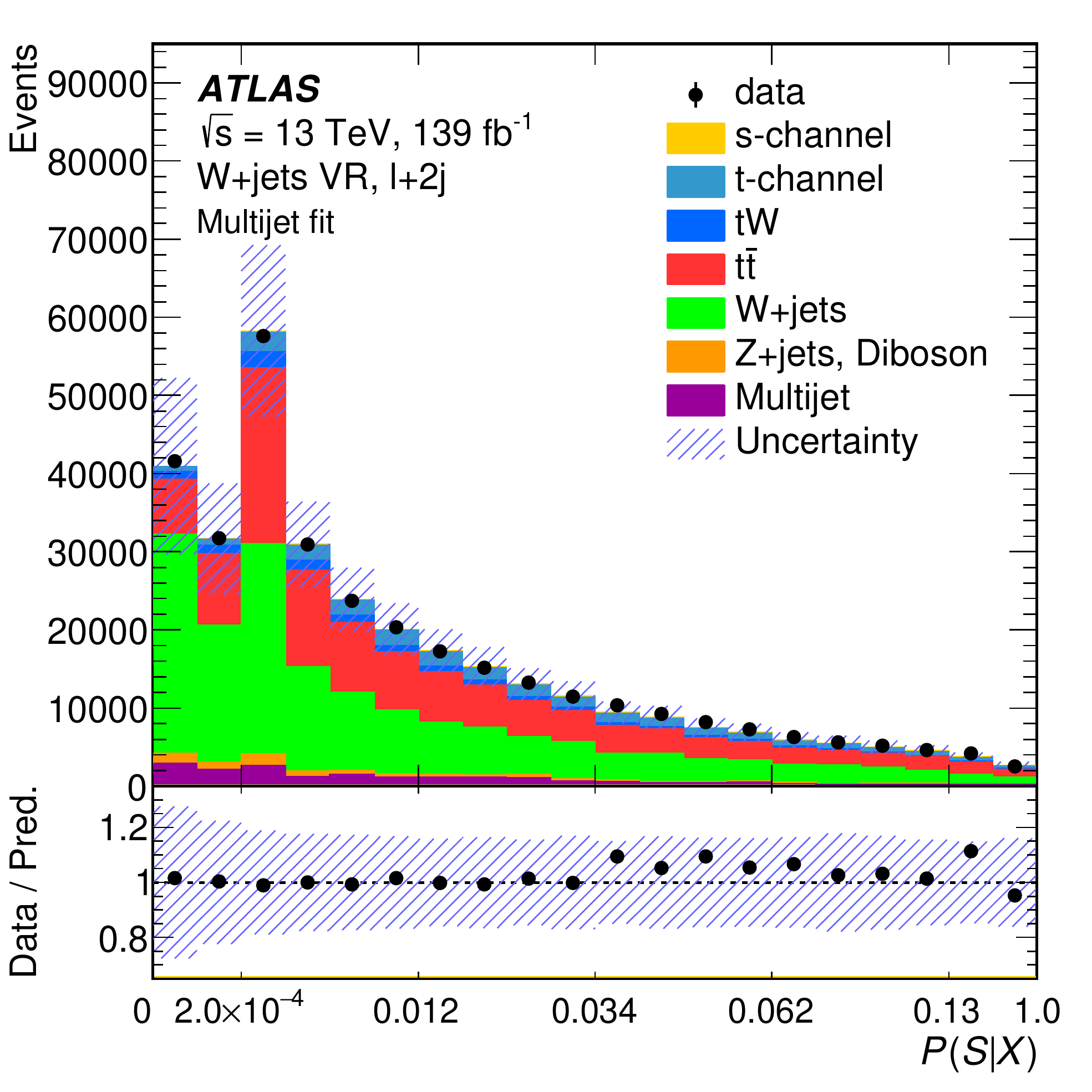}
\label{subfig:mem-disc-ctrl-wjets}
}\\
\subfigure[]{
\includegraphics[width=0.4\textwidth]{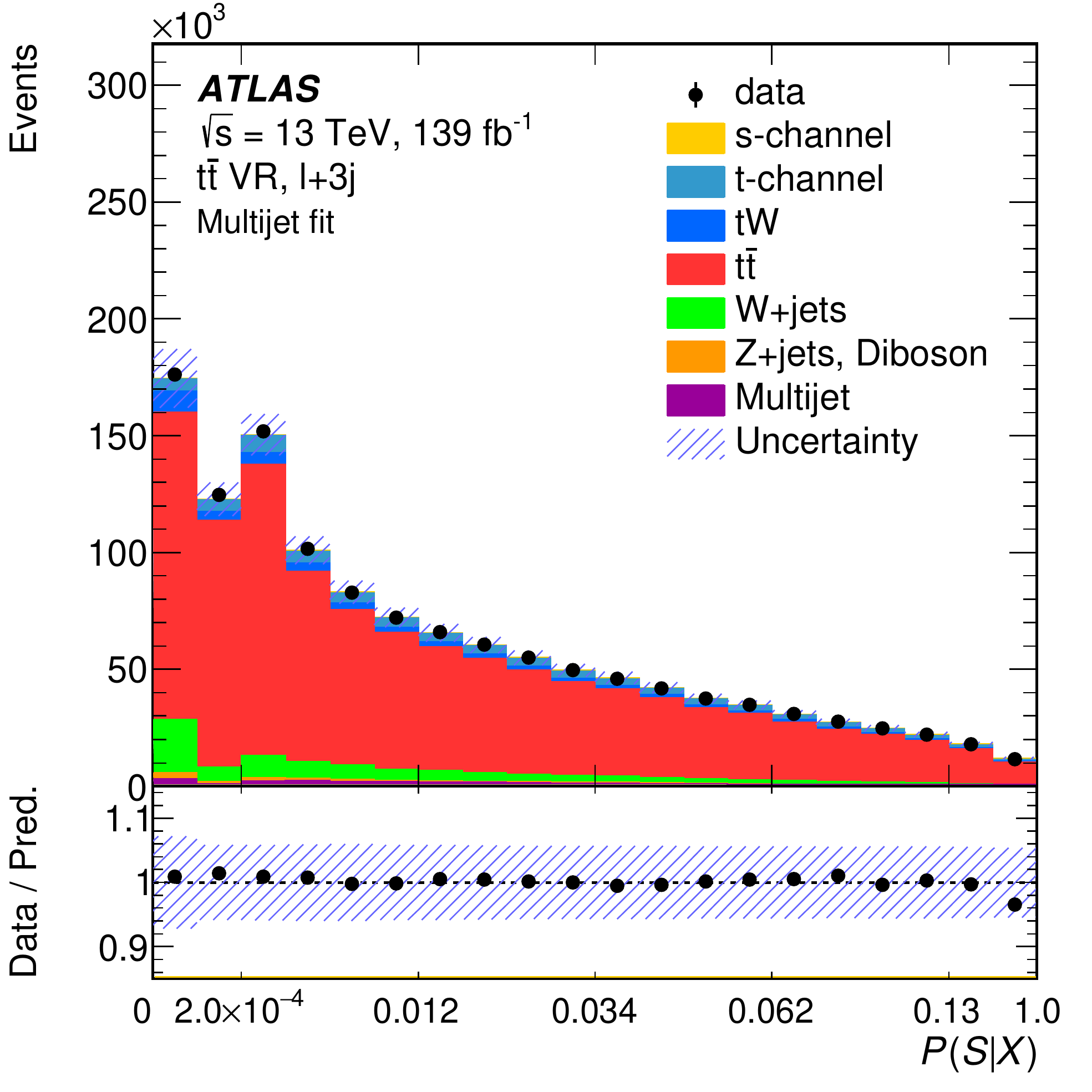}
\label{subfig:mem-disc-ctrl-ttbar3j}
}
\subfigure[]{
\includegraphics[width=0.4\textwidth]{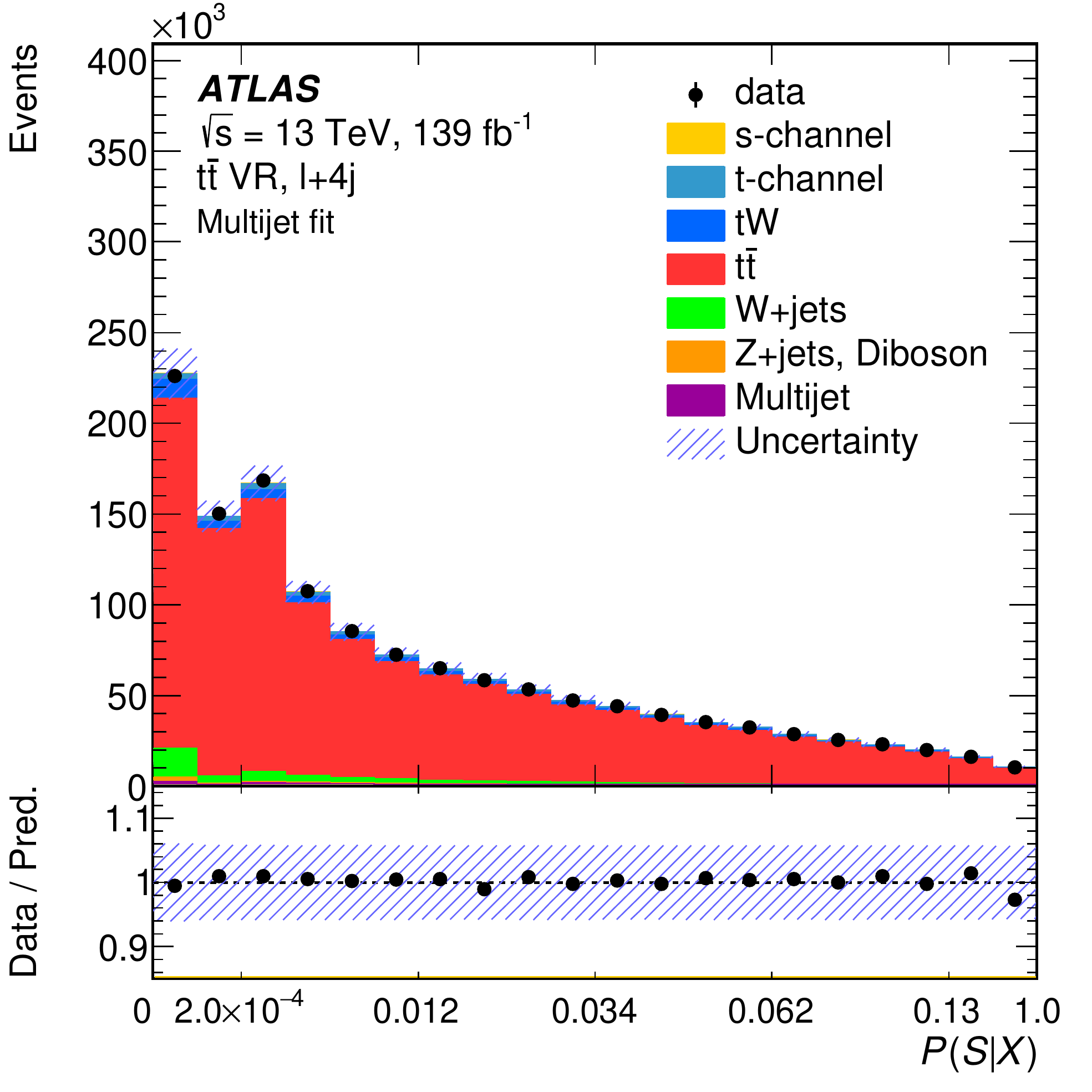}
\label{subfig:mem-disc-ctrl-ttbar4j}
}
\caption{
Distributions of the MEM discriminant $P(S|X)$ in the \subref{subfig:mem-disc-ctrl-wjets}~\WVR,
\subref{subfig:mem-disc-ctrl-ttbar3j}~\ttVR[3], and \subref{subfig:mem-disc-ctrl-ttbar4j}~\ttVR[4].
Simulated events are normalised to the expected number of events given the integrated luminosity, after applying the normalisation factors obtained in the multijet fit presented in Section~\ref{sec:backgrounds}. The uncertainty bands indicate the simulation's statistical uncertainty and the normalisation uncertainties for the various processes in each bin, summed in quadrature.
The ratio of the observed number to the predicted number of events in each bin is shown in the lower panel, with different vertical axis ranges.
The binning is the same as the optimised binning used in the signal extraction fit
described in Section~\ref{sec:signal_fit}, resulting in a non-linear horizontal scale.
}
\label{fig:mem-disc-validation}
\end{figure}

\clearpage


\section{Systematic uncertainties}
\label{sec:systematics}

Systematic uncertainties affect the signal acceptance and the background normalisations.
Furthermore, most of them affect the shape of the MEM discriminant distribution
because they affect the four-momenta of the reconstructed objects, or their efficiencies.
Uncertainties related to the detector and to the modelling of the signal and background processes are taken into account.
In the statistical analysis, an independent nuisance parameter is assigned to each source of systematic uncertainty.
Several of the systematic uncertainties are decomposed into components from multiple independent sources.
 
The uncertainty in the combined 2015--2018 integrated luminosity is 1.7\%~\cite{ATLAS-CONF-2019-021},
obtained using the LUCID-2 detector~\cite{Avoni:2018iuv} for the primary luminosity measurement.
An uncertainty associated with the modelling of pile-up in the simulation is included
to cover the difference between the predicted and measured inelastic cross-section values~\cite{STDM-2015-05}.
 
The jet energy scale uncertainty is derived by combining information from test-beam data,
LHC collision data and simulation, and the jet energy resolution uncertainty
is obtained by combining dijet \pt-balance measurements and simulation~\cite{JETM-2018-05}.
Additional considerations related to jet flavour, pile-up corrections,
$\eta$ dependence and high-\pt jets are included.
A total of 31 independent contributions are considered for the jet energy scale uncertainty,
and 9 for the jet energy resolution uncertainty.
The efficiency of the JVT algorithm to identify and remove jets from pile-up is measured with $Z \to \mu^+\mu^-$
events in data using techniques similar to those used in Ref.~\cite{PERF-2014-03}.
 
The efficiency to correctly tag $b$-jets is measured using dileptonic \ttbar\ events~\cite{FTAG-2018-01}.
The mis-tag rate for $c$-jets is measured using semileptonic \ttbar\ events,
exploiting the $c$-jets from the hadronic \Wboson-boson decays
using techniques similar to those in Ref.~\cite{ATLAS-CONF-2018-001}.
The mis-tag rate for light-jets is measured using a negative-tag method,
similar to that in Ref.~\cite{ATLAS-CONF-2018-006}, applied to \Zjets events.
Uncertainties on this efficiency and these mis-tag rates are due to reconstructed object calibrations
and to the modelling of the different processes, and are decomposed into sets of uncorrelated sources
of uncertainty: 45, 20, and 20 components for $b$-, $c$- and light-jets, respectively.
 
Uncertainties associated with leptons arise from the trigger, reconstruction,
identification, and isolation, as well as the lepton momentum scale and resolution.
Efficiencies are measured, and calibrations of the scale and resolution are performed,
using leptons in $Z\to \ell^+\ell^-$ and $J/\psi \to \ell^+\ell^-$ events~\cite{EGAM-2018-01,PERF-2015-10}.
Systematic uncertainties in these measurements account for 22 independent sources.
 
All uncertainties related to the energy scales or resolution
of the reconstructed objects are propagated to the calculation of the missing transverse momentum.
Three additional uncertainties associated with the scale and resolution of the soft term are also included.
 
The modelling of the cross-section and acceptance of the two dominant backgrounds, \ttbar and \Wjets production,
is affected by several systematic uncertainties.
Therefore, the normalisations of these processes are measured in the signal extraction fit (cf.\ Section~\ref{sec:signal_fit}),
and no constraints are placed on these two freely floating parameters.
For single-top \tchan and \tW production, uncertainties in the theoretical
cross-sections of \SI{4}{\%} and \SI{5}{\%}, respectively, are propagated to the statistical analysis.
For $Z$+jets and diboson processes, an uncertainty of \SI{60}{\%} is considered, including $\pm5$\%
for the inclusive cross-section summed in quadrature with $\pm24$\%
per additional jet and $\pm50$\% for the heavy-flavour component~\cite{Ellis:1985vn,Berends:1990ax,TOPQ-2015-01}.
Finally, an uncertainty of \SI{30}{\%} is assigned to the normalisation of the multijet background,
motivated by the change in the multijet normalisation when altering
various settings in the multijet fit described in Section~\ref{sec:backgrounds},
or by comparisons with other methods of estimating this background,
as used in previous analyses~\cite{ATLAS-CONF-2014-058,TOPQ-2015-05}.
 
Uncertainties due to the modelling of the different processes are taken into account.
For each of the single-top \schan, \tchan\ and \tW\ processes, and the \ttbar process,
four sources related to the modelling of initial- and final-state radiation (ISR/FSR) are considered:
two independent sources estimated by separately varying the renormalisation and factorisation scales in the ME generator by factors of 0.5 and 2,
one source estimated by varying the renormalisation scale for QCD emission in the initial-state radiation in \PYTHIA[8]
(corresponding to the Var3c variation of the A14 tune~\cite{ATL-PHYS-PUB-2014-021}),
and one source estimated by halving and doubling the renormalisation scale
for QCD emission in the final-state radiation in \PYTHIA[8].
One additional source is also estimated for each of these four top-quark processes to account for the parton shower and hadronisation model,
by comparing the nominal \POWHEGBOX\!+\,\PYTHIA[8] MC sample with a \POWHEGBOX\!+\,\HERWIG[7.04] sample.
The uncertainty associated with the algorithm removing the overlap between \tW and \ttbar\ production at NLO~\cite{Frixione:2008yi} is assessed
by comparing the nominal \POWHEGBOX\!+\,\PYTHIA[8] \tW sample produced using the diagram removal scheme
with an alternative sample produced with the same generator but using the diagram subtraction scheme.
For the dominant \ttbar background, two additional sources are considered:
one source related to soft gluon resummation is estimated by setting the $h_{\text{damp}}$ parameter
to $3.0\cdot\mtop$, instead of the nominal $1.5\cdot\mtop$, in the \POWHEGBOX\!+\,\PYTHIA[8] \ttbar MC sample,
and one source related to the matching of the ME to the parton shower is estimated
by comparing the nominal \POWHEGBOX\!+\,\PYTHIA[8] \ttbar MC sample with a \MGNLO\!+\,\PYTHIA[8] sample.
One source of uncertainty related to the shape of the \Wjets background is evaluated by using the envelope of the independent variations
of the renormalisation and factorisation scales in the ME of the \Wjets MC sample.
Two sources of uncertainty affecting the shape of the multijet background are considered, one for the electron channel and one for the muon channel.
These two sources are estimated by varying the criteria related to the fraction of energy deposited in the electromagnetic calorimeter
for the jet-electron model, and to the tracking isolation in the anti-muon model, respectively (cf.\ Section~\ref{sec:backgrounds}).
All these modelling uncertainties may affect both the shape of the MEM discriminant and the normalisation of each process,
except for \ttbar\ and \Wjets, for which the normalisation is extracted in the fit.
 
The choice of PDFs affects the modelling of top-quark processes, and these uncertainties are evaluated using the \PDFforLHC[15] combined PDF error sets~\cite{Butterworth:2015oua},
which contain 30 symmetric eigenvectors.
Four independent sets of 30 uncertainty sources are considered, one for each process.
For single-top \tchan production, the $\PDFforLHC[15]$\_nlo\_nf4\_30 set is used,
while for single-top \schan and \tW\ production, and \ttbar production, the $\PDFforLHC[15]$\_nlo\_30 set is used.


\section{Signal extraction}
\label{sec:signal_fit}

The single-top-quark \schan production cross-section is measured by means of a binned profile maximum-likelihood
fit of the MEM discriminant in the signal region.
The binning of the MEM discriminant is optimised in the signal region using the procedure described in Ref.~\cite{HIGG-2013-23},
in order to maximise the expected sensitivity while keeping the total statistical uncertainty of the predicted number of events
in each bin at a level adjusted to avoid biases due to fluctuations.
This results in bins of unequal width, with wider bins in regions with a large
signal contribution, while preserving a sufficiently large number of background events in each bin.
In this paper, MEM discriminant distributions are presented with a non-linear horizontal scale,
in such a way that the histogram bins appear to have a constant width.
Values of $P(S|X)$ lower than $2.0\times10^{-4}$ are not taken into account for the signal
extraction because of the very low signal-to-background ratio in this range;
this rejects 21\% (18\%) of the expected \ttbar (\Wjets) background events,
while the expected signal yield is reduced by less than 1\%.
The electron and muon channels, which have similar sensitivity to the signal,
are merged regardless of the lepton charge or flavour or of the various data-taking periods of the LHC,
while preserving correlations of the later described nuisance parameters associated to systematic uncertainties,
in order to measure the combined production cross-section of single top quarks and top antiquarks.
 
The binned likelihood function ${\cal L}(\mu,\theta)$ used in the fit consists of a product of Poisson probability terms over the signal region histogram bins.
The likelihood function depends on the signal strength $\mu$, defined as $\mu=\sigma/\sigma^{\mathrm{SM}}$, where
$\sigma$ is the observed signal production cross-section,
and on a set of nuisance parameters $\theta$, which characterise the effects of systematic uncertainties in the signal and background predictions.
They are implemented in the likelihood function as Gaussian priors,
with the exception of the unconstrained normalisation factors for the \ttbar and \Wjets backgrounds.
The statistical uncertainty in the predictions, which incorporates the statistical
uncertainty arising from the limited number of simulated events, is included in the likelihood function
in the form of additional nuisance parameters, one for each of the considered bins, and constrained by Poisson priors.
The test statistic $t_{\mu}$ is defined as the profile likelihood ratio:
$t_{\mu}=-2\ln({\cal L}(\mu,\hat{\hat{\theta}}_\mu)/{\cal L}(\hat{\mu},\hat{\theta}))$,
where $\hat{\mu}$ and $\hat{\theta}$ are the values of the parameters which maximise the likelihood function,
and $\hat{\hat{\theta}}_\mu$ are the values of the nuisance parameters
which maximise the likelihood function for a given value of $\mu$~\cite{Cowan:2010js}.
This test statistic, implemented in a framework based on RooStats~\cite{RooFit,RooStat} and HistFactory~\cite{Cranmer:1456844},
is used to assess the compatibility of the observed data with the background-only hypothesis, and to make statistical inferences about $\mu$.


\section{Results}

A comparison of the MEM discriminant distributions in data and simulation
is shown in Figure~\ref{fig:mem-disc-SR}, before and after the fit to data.
The uncertainty in the total prediction before the fit in Figure~\ref{subfig:mem-disc-SR-prefit}
is dominated by signal modelling, the jet energy resolution and scale,
ISR/FSR in top-quark processes, and \ttbar modelling uncertainties.
The distribution after the fit in Figure~\ref{subfig:mem-disc-SR-postfit}
shows good agreement within the total uncertainty band,
and the global goodness of fit calculated using the saturated model~\cite{Cousins:2013} is $10.3\%$.
The measured distribution of the MEM discriminant for the signal contribution is shown in
Figure~\ref{fig:mem-disc-SR-DataMinusBkg}, and is compared with the data after the subtraction
of all background contributions.
The expected event yields for the different processes, before and after the fit,
are shown in Table~\ref{tab:Yields}, together with the observed event yield in data.
The measured \ttbar and \Wjets normalisation factors are
$0.81^{+0.13}_{-0.12}$ and $1.37^{+0.35}_{-0.31}$, respectively;
their correlation is 51\%, and their respective correlations with the signal cross-section are 55\% and 27\%.
After the fit, none of the other nuisance parameters is pulled by more than 0.8 standard deviation,
constrained to less than 74\% of its pre-fit impact on the cross-section by the fit,
or has a correlation with the fitted cross-section larger than 33\%.
 
Several tests were performed in order to assess the robustness of this fit, called 'baseline fit' in this paragraph.
First, instead of being left unconstrained, the \ttbar and \Wjets normalisations were assigned
an uncertainty on their cross-sections.
In such case, the nuisance parameters associated to these uncertainties
were pulled in the same directions as the normalisation factors in the baseline fit.
Second, a fit of the background-only hypothesis to the data was performed, keeping only bins of the discriminant distribution
in which the expected signal-to-background ratio does not exceed 5\%.
In such a fit, the \ttbar and \Wjets normalisation factors were pulled by values
compatible to those obtained in the baseline fit.
Furthermore, the post-fit nuisance parameters of this background-only fit were propagated to the signal and background MC samples,
with signal cross-section set to its prediction from theory, in order to build a 'realistic' pseudo-data set.
A signal-plus-background fit to this pseudo-data set was then performed, and the measured cross-section
was compatible with the predicted cross-section.
Finally, the linearity of the statistical model was checked, by performing fits to pseudo-data sets produced from
the nominal predictions (Asimov datasets), where the expected signal cross-section was increased or decreased.
In such fits, the measured and expected cross-sections were in agreement.

\begin{figure}[!htb]
\centering
\subfigure[]{
\includegraphics[width=0.4\textwidth]{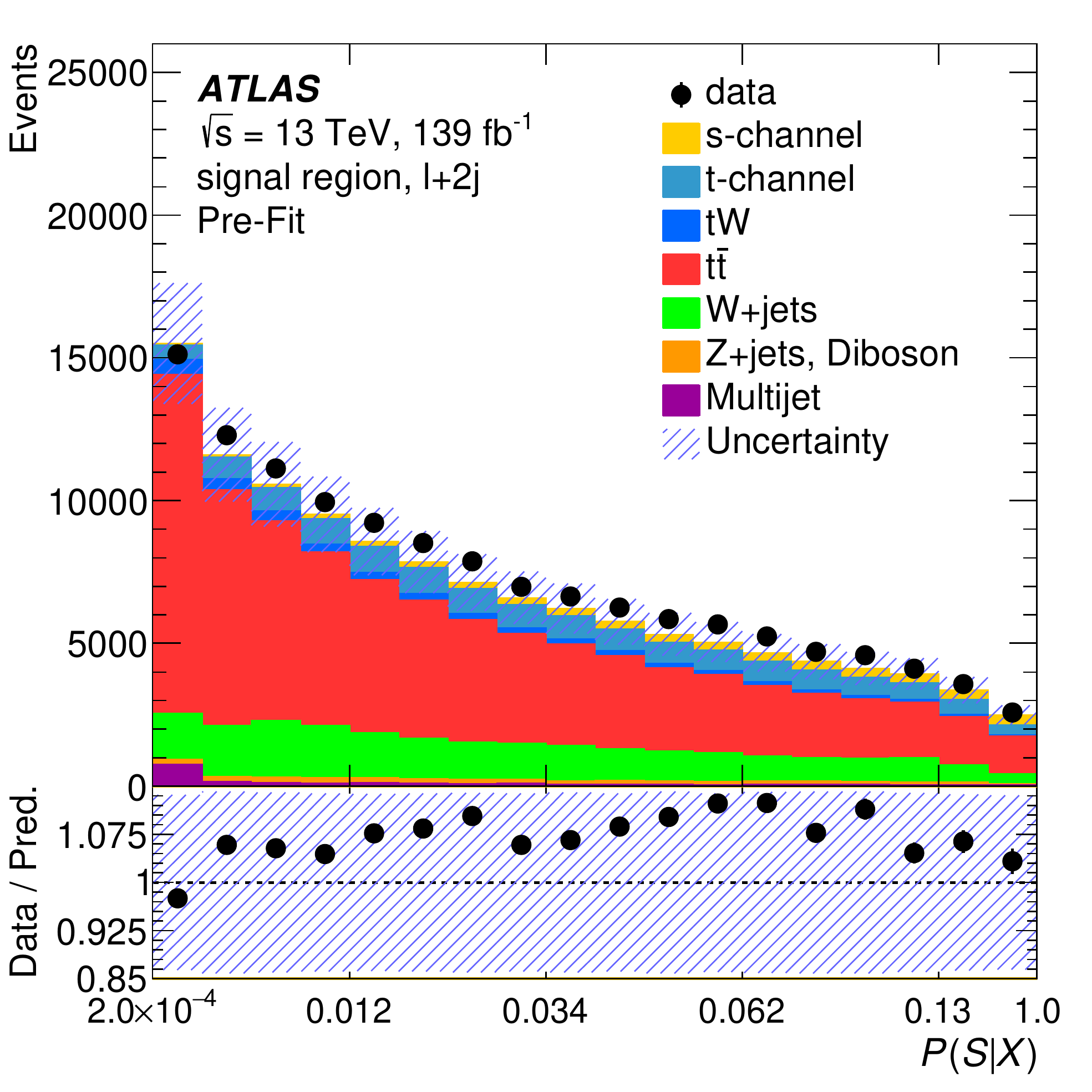}
\label{subfig:mem-disc-SR-prefit}
}
\subfigure[]{
\includegraphics[width=0.4\textwidth]{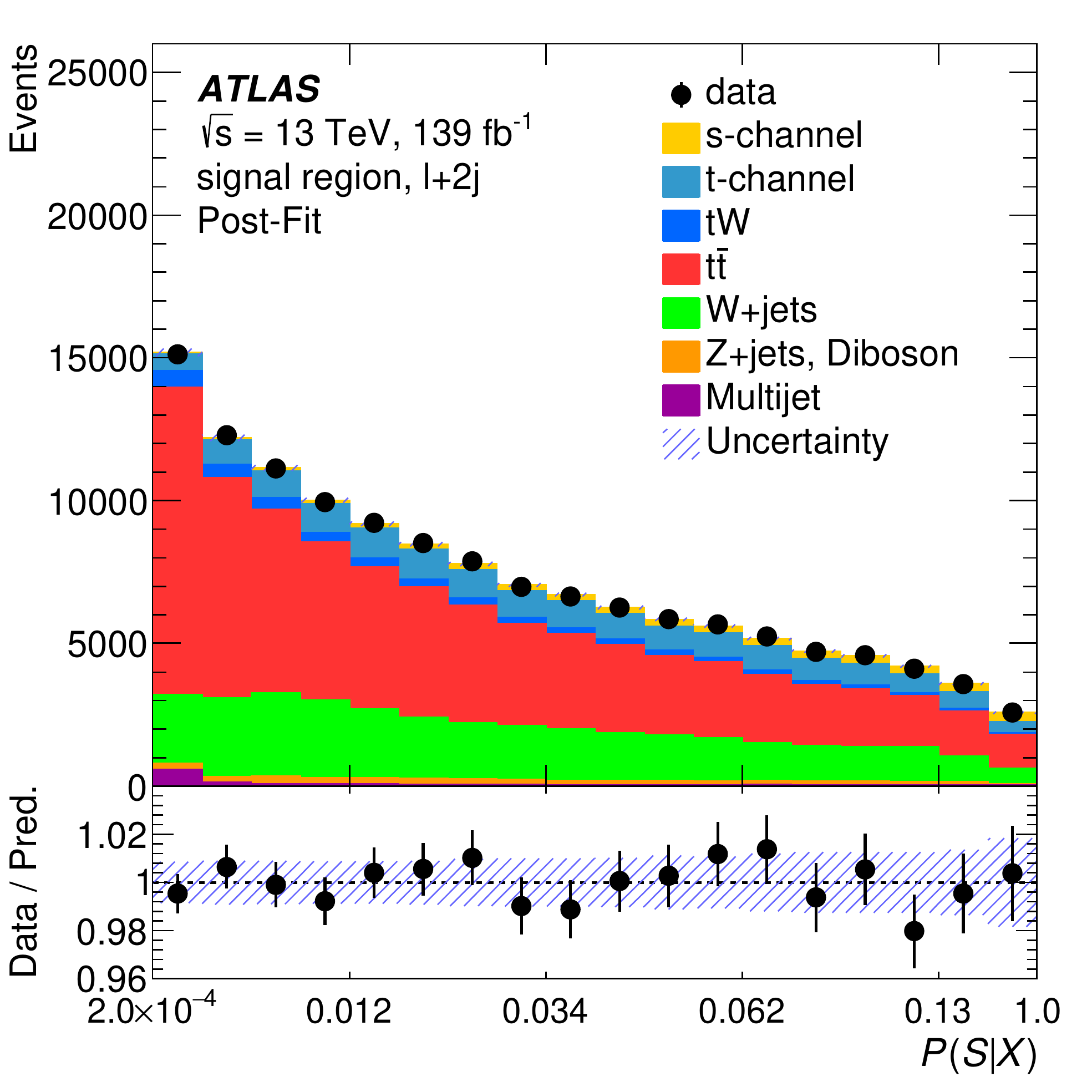}
\label{subfig:mem-disc-SR-postfit}
}
\caption{
Distributions of the MEM discriminant $P(S|X)$ in the SR
\subref{subfig:mem-disc-SR-prefit} before and \subref{subfig:mem-disc-SR-postfit} after the fit to data,
for MEM discriminant values larger than $2.0\times10^{-4}$.
The lower panels show the ratio of the data to the prediction, with different vertical axis ranges.
The uncertainty bands indicate the total uncertainties and their correlations in each bin.
The uncertainties in the \ttbar and \Wjets normalisation factors, as well as in the \schan signal cross-section,
are not defined pre-fit and therefore only included in the post-fit uncertainties in~\subref{subfig:mem-disc-SR-postfit}.
The binning is the same as the optimised binning used in the fit, resulting in a non-linear horizontal scale.
}
\label{fig:mem-disc-SR}
\end{figure}
 
\begin{figure}[!htb]
\centering
\includegraphics[width=0.4\textwidth]{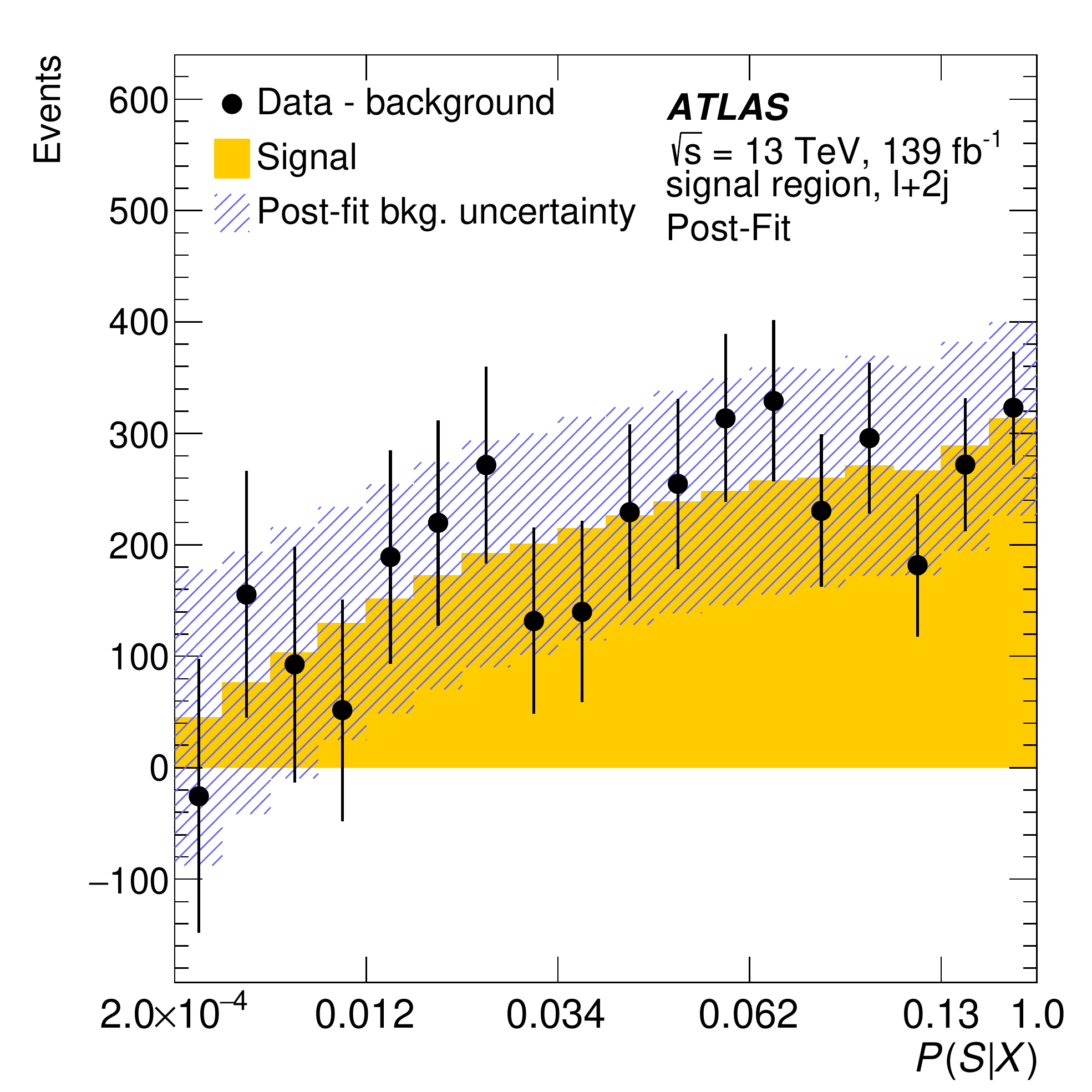}
\caption{
Distribution of the MEM discriminant $P(S|X)$ in the SR after the fit to data,
for MEM discriminant values larger than $2.0\times10^{-4}$,
after subtraction of all backgrounds.
The fitted distribution for the simulation of the signal is shown together with
the post-fit uncertainty in the backgrounds.
The binning is the same as the optimised binning used in the fit, resulting in a non-linear horizontal scale.
}
\label{fig:mem-disc-SR-DataMinusBkg}
\end{figure}

\clearpage
 
\begin{table}[!htb]\centering
\sisetup{group-minimum-digits=4}
\begin{tabular}{l|
r@{$\,\pm\,$}l
r@{$\,\pm\,$}l}
\toprule
\midrule
\multirow{2}{*}{Process} & \multicolumn{4}{c}{Event yield} \\
& \multicolumn{2}{c}{Pre-fit} & \multicolumn{2}{c}{Post-fit} \\
\midrule\midrule
\schan       & \numRF{4198.07}{3} & \numRF{711.687}{2}     & \numRF{3652.86}{2} & \numRF{1130.57}{2} \\
\tchan       & \numRF{13004.7}{3} & \numRF{1980.94}{2}     & \numRF{14970.8}{3} & \numRF{2279.02}{2} \\
\tW          & \numRF{3676.49}{3} & \numRF{972.375}{2}     & \numRF{4249.54}{3} & \numRF{1123.36}{2} \\
\ttbar       & \numRF{75789.7}{2} & \numRF{11587.1}{2}     & \numRF{70627.3}{3} & \numRF{4189.24}{2} \\
\Wjets       & \numRF{21465.4}{3} & \numRF{2932.32}{2}     & \numRF{32228.4}{3} & \numRF{5037.21}{2} \\
\Zjets, \VV  & \numRF{2367.18}{2} & \numRF{1444.84}{2}     & \numRF{2882.31}{2} & \numRF{1618.07}{2} \\
Multijet     & \numRF{2152.41}{3} & \numRF{647.959}{2}     & \numRF{1695.44}{3} & \numRF{542.951}{2} \\
\midrule
Total        & \numRF{122654}{3} & \numRF{16856.5}{2}      & \numRF{130307}{5}  & \numRF{618.162}{2} \\
\midrule\midrule
Data         & \multicolumn{4}{c}{\num{130310}} \\
\bottomrule\bottomrule
\end{tabular}
\caption{
Pre-fit and post-fit event yields in the SR, for MEM discriminant values larger than $2.0\times10^{-4}$.
The central value of the event yield for each process is calculated by summing the values of the discriminant bin contents,
using the nominal expected yield for the pre-fit value, and the best-fit estimate for the post-fit value.
The error includes statistical and systematic uncertainties summed in quadrature.
All sources of systematic uncertainties are included, taking into account correlations and anti-correlations in the post-fit case.
The uncertainties in the \ttbar and \Wjets normalisation factors, as well as in the \schan signal cross-section,
are not defined pre-fit and therefore only included in the post-fit uncertainties.
}
\label{tab:Yields}
\end{table}

The measured cross-section is $\sigma=8.2\pm0.6\mathrm{~(stat.)}^{+3.4}_{-2.8}\mathrm{~(syst.)}\,\mathrm{pb}=8.2^{+3.5}_{-2.9}\mathrm{~(total)}\,\mathrm{pb}$,
which is compatible with the SM prediction of $\sigma^{\mathrm{SM}}=10.32^{+0.40}_{-0.36}\,\mathrm{pb}$.
A summary of the main sources of uncertainty is presented in Table~\ref{tab:systematics}.
The largest contribution arises from the \ttbar normalisation.
Uncertainties in the jet energy scale and resolution, and in the signal generator modelling, also play an important role,
followed by the modelling of ISR/FSR in top-quark processes, the statistical uncertainty of the MC predictions,
the \ttbar\ generator modelling uncertainties, and the flavour-tagging uncertainties.
The observed signal significance is 3.3~standard deviations above the background-only hypothesis,
while the expected significance is 3.9~standard deviations.
 
\clearpage
 
\begin{table}[!htb]\centering
 
\begin{tabular}{l|l|r@{/}l}
\toprule\toprule
\multicolumn{2}{l|}{Source}                & \multicolumn{2}{c}{$\Delta\sigma/\sigma$ $[\%]$} \\
\midrule\midrule
\multicolumn{2}{l|}{\ttbar normalisation}  & $+\numRF{24.131087}{2}$&$-\numRF{16.885713}{2}$ \\
 
\multicolumn{2}{l|}{\ttbar shape modelling} & $+\numRF{17.732814}{2}$&$-\numRF{14.960322}{2}$ \\%
& ISR/FSR           & $+\numRF{12.560875}{2}$&$-\numRF{10.680888}{2}$ \\
& PS \& had.        & $+\numRF{11.602321}{2}$&$-\numRF{9.5116510}{1}$ \\
& ME/PS matching    & $+\numRF{9.5630149}{1}$&$-\numRF{8.2135241}{1}$ \\
& $h_{\text{damp}}$ & \multicolumn{2}{c}{$<1$~~~~} \\
 
\multicolumn{2}{l|}{\schan modelling} & $+\numRF{18.047462}{2}$&$-\numRF{8.4941193}{1}$ \\
& PS \& had. & $+\numRF{17.806173}{2}$&$-\numRF{8.3966692}{1}$ \\
& ISR/FSR    & $+\numRF{2.8619625}{1}$&$-\numRF{1.4835952}{1}$ \\
 
\multicolumn{2}{l|}{Jet energy resolution} & $+\numRF{18.445697}{2}$&$-\numRF{12.303184}{2}$  \\
\multicolumn{2}{l|}{Jet energy scale} & $+\numRF{18.095316}{2}$&$-\numRF{13.183922}{2}$ \\

\multicolumn{2}{l|}{MC statistics} & $+\numRF{12.925208}{2}$&$-\numRF{11.200033}{2}$  \\
\multicolumn{2}{l|}{Flavour tagging} & $+\numRF{11.802865}{2}$&$-\numRF{9.6992910}{1}$ \\
\multicolumn{2}{l|}{\Wjets normalisation} & $+\numRF{11.099338}{2}$&$-\numRF{7.5381410}{1}$ \\
 
\multicolumn{2}{l|}{PDFs} & $+\numRF{10.145051}{2}$&$-\numRF{8.6508753}{1}$ \\
& \ttbar & $+\numRF{10.080771}{2}$&$-\numRF{8.5876300}{1}$ \\
& \schan & \multicolumn{2}{c}{$\pm 1$~~~~} \\
& \tchan & \multicolumn{2}{c}{$\pm 1$~~~~} \\
& \tW    & \multicolumn{2}{c}{$\pm 1$~~~~} \\
 
\multicolumn{2}{l|}{\tchan modelling} &  \multicolumn{2}{c}{$\pm 6$~~~~} \\
& PS \& had. & \multicolumn{2}{c}{$\pm 5$~~~~} \\
& ISR/FSR    & \multicolumn{2}{c}{$\pm 4$~~~~} \\
 
\multicolumn{2}{l|}{\Wjets $\muR/\muF$ shape} & $+\numRF{5.8878023}{1}$&$-\numRF{5.1424188}{1}$ \\
\multicolumn{2}{l|}{Normalisation of other processes} & $+\numRF{5.7854279}{1}$&$-\numRF{5.2986698}{1}$ \\
\multicolumn{2}{l|}{Pile-up} & $+\numRF{4.8487680}{1}$&$-\numRF{3.2733540}{1}$ \\
\multicolumn{2}{l|}{Luminosity} & $+\numRF{3.7609202}{1}$&$-\numRF{2.6761533}{1}$ \\
 
\multicolumn{2}{l|}{\tW modelling}  & $+\numRF{1.2714979}{1}$&$-\numRF{1.8057690}{1}$ \\
& PS \& had.      & \multicolumn{2}{c}{$\pm 1$~~~~} \\
& \ttbar\ overlap & \multicolumn{2}{c}{$\pm 1$~~~~} \\
& ISR/FSR         & \multicolumn{2}{c}{$\pm 1$~~~~} \\
 
\multicolumn{2}{l|}{Missing transverse momentum} & \multicolumn{2}{c}{$\pm 1$~~~~} \\
\multicolumn{2}{l|}{Multijet shape modelling} & \multicolumn{2}{c}{$\pm 1$~~~~} \\
\multicolumn{2}{l|}{Other detector sources} & \multicolumn{2}{c}{$\pm 1$~~~~} \\
\midrule
\multicolumn{2}{l|}{Systematic uncertainties} & $+\numRF{41.714152}{2}$&$-\numRF{34.149837}{2}$ \\
\midrule
\multicolumn{2}{l|}{Statistical uncertainty} & \multicolumn{2}{c}{$\pm 8$~~~~} \\
\midrule\midrule
\multicolumn{2}{l|}{Total} & $+\numRF{42.381076}{2}$&$-\numRF{34.974337}{2}$ \\
\bottomrule\bottomrule
 
\end{tabular}
\caption{
Observed impact of the different sources of uncertainty on the measured \schan signal cross-section, grouped by categories.
The impact of each category is obtained by repeating the fit
after having fixed the set of nuisance parameters corresponding to that category,
subtracting the square of the resulting uncertainty from the square of the uncertainty found in the full fit,
and calculating the square root.
The statistical uncertainty is obtained by repeating the fit
after having fixed all nuisance parameters, including the \ttbar\ and \Wjets\ normalisation factors.
'PS \& had.' refers to the parton shower and hadronisation model,
'ME/PS matching' to the matching of the ME to the parton shower,
and '\ttbar\ overlap' to the algorithm removing the overlap between \tW and \ttbar\ production at NLO,
as described in Section~\ref{sec:systematics}.
}
\label{tab:systematics}
\end{table}


\FloatBarrier

\section{Conclusion}

A measurement of the \schan single-top-quark production cross-section in \pp collisions
at a centre-of-mass energy of \SI{13}{\TeV} recorded by the ATLAS detector at the LHC is presented.
The analysis was performed using data collected in 2015--2018 corresponding to an integrated luminosity of \SI{139}{\ifb}.
The selected events contain one electron or muon, two jets which are both $b$-tagged, and missing transverse momentum.
The matrix element method was used to obtain a discriminant to separate the signal
from the background events, and the signal was extracted using a profile likelihood fit.
The measured signal cross-section, combining top-quark and top-antiquark production, is
$\sigma=8.2^{+3.5}_{-2.9}\,\SI{}{pb}$,
in agreement with the SM prediction of
$\sigma^{\mathrm{SM}}=10.3\pm0.4\,\SI{}{pb}$.
The observed (expected) signal significance is 3.3 (3.9) standard deviations
above the background-only hypothesis.
This result provides a measurement of the cross-section of this process
at a new centre-of-mass energy, which allows
the energy dependence of this observable to be probed.


\section*{Acknowledgements}


We thank CERN for the very successful operation of the LHC, as well as the
support staff from our institutions without whom ATLAS could not be
operated efficiently.
 
We acknowledge the support of
ANPCyT, Argentina;
YerPhI, Armenia;
ARC, Australia;
BMWFW and FWF, Austria;
ANAS, Azerbaijan;
CNPq and FAPESP, Brazil;
NSERC, NRC and CFI, Canada;
CERN;
ANID, Chile;
CAS, MOST and NSFC, China;
Minciencias, Colombia;
MEYS CR, Czech Republic;
DNRF and DNSRC, Denmark;
IN2P3-CNRS and CEA-DRF/IRFU, France;
SRNSFG, Georgia;
BMBF, HGF and MPG, Germany;
GSRI, Greece;
RGC and Hong Kong SAR, China;
ISF and Benoziyo Center, Israel;
INFN, Italy;
MEXT and JSPS, Japan;
CNRST, Morocco;
NWO, Netherlands;
RCN, Norway;
MEiN, Poland;
FCT, Portugal;
MNE/IFA, Romania;
MESTD, Serbia;
MSSR, Slovakia;
ARRS and MIZ\v{S}, Slovenia;
DSI/NRF, South Africa;
MICINN, Spain;
SRC and Wallenberg Foundation, Sweden;
SERI, SNSF and Cantons of Bern and Geneva, Switzerland;
MOST, Taiwan;
TENMAK, T\"urkiye;
STFC, United Kingdom;
DOE and NSF, United States of America.
In addition, individual groups and members have received support from
BCKDF, CANARIE, Compute Canada and CRC, Canada;
PRIMUS 21/SCI/017 and UNCE SCI/013, Czech Republic;
COST, ERC, ERDF, Horizon 2020 and Marie Sk{\l}odowska-Curie Actions, European Union;
Investissements d'Avenir Labex, Investissements d'Avenir Idex and ANR, France;
DFG and AvH Foundation, Germany;
Herakleitos, Thales and Aristeia programmes co-financed by EU-ESF and the Greek NSRF, Greece;
BSF-NSF and MINERVA, Israel;
Norwegian Financial Mechanism 2014-2021, Norway;
NCN and NAWA, Poland;
La Caixa Banking Foundation, CERCA Programme Generalitat de Catalunya and PROMETEO and GenT Programmes Generalitat Valenciana, Spain;
G\"{o}ran Gustafssons Stiftelse, Sweden;
The Royal Society and Leverhulme Trust, United Kingdom.
 
The crucial computing support from all WLCG partners is acknowledged gratefully, in particular from CERN, the ATLAS Tier-1 facilities at TRIUMF (Canada), NDGF (Denmark, Norway, Sweden), CC-IN2P3 (France), KIT/GridKA (Germany), INFN-CNAF (Italy), NL-T1 (Netherlands), PIC (Spain), ASGC (Taiwan), RAL (UK) and BNL (USA), the Tier-2 facilities worldwide and large non-WLCG resource providers. Major contributors of computing resources are listed in Ref.~\cite{ATL-SOFT-PUB-2021-003}.


\printbibliography

\clearpage
 
 
\begin{flushleft}
\hypersetup{urlcolor=black}
{\Large The ATLAS Collaboration}

\bigskip

\AtlasOrcid[0000-0002-6665-4934]{G.~Aad}$^\textrm{\scriptsize 101}$,
\AtlasOrcid[0000-0002-5888-2734]{B.~Abbott}$^\textrm{\scriptsize 119}$,
\AtlasOrcid[0000-0002-7248-3203]{D.C.~Abbott}$^\textrm{\scriptsize 102}$,
\AtlasOrcid[0000-0002-1002-1652]{K.~Abeling}$^\textrm{\scriptsize 55}$,
\AtlasOrcid[0000-0002-8496-9294]{S.H.~Abidi}$^\textrm{\scriptsize 29}$,
\AtlasOrcid[0000-0002-9987-2292]{A.~Aboulhorma}$^\textrm{\scriptsize 35e}$,
\AtlasOrcid[0000-0001-5329-6640]{H.~Abramowicz}$^\textrm{\scriptsize 150}$,
\AtlasOrcid[0000-0002-1599-2896]{H.~Abreu}$^\textrm{\scriptsize 149}$,
\AtlasOrcid[0000-0003-0403-3697]{Y.~Abulaiti}$^\textrm{\scriptsize 116}$,
\AtlasOrcid[0000-0003-0762-7204]{A.C.~Abusleme~Hoffman}$^\textrm{\scriptsize 136a}$
% ...
\bigskip
\\

% ...
$^{29}$Physics Department, Brookhaven National Laboratory, Upton NY; United States of America.\\
% ...
$^{35}$$^{(a)}$Facult\'e des Sciences Ain Chock, R\'eseau Universitaire de Physique des Hautes Energies - Universit\'e Hassan II, Casablanca;$^{(b)}$Facult\'{e} des Sciences, Universit\'{e} Ibn-Tofail, K\'{e}nitra;$^{(c)}$Facult\'e des Sciences Semlalia, Universit\'e Cadi Ayyad, LPHEA-Marrakech;$^{(d)}$LPMR, Facult\'e des Sciences, Universit\'e Mohamed Premier, Oujda;$^{(e)}$Facult\'e des sciences, Universit\'e Mohammed V, Rabat;$^{(f)}$Institute of Applied Physics, Mohammed VI Polytechnic University, Ben Guerir; Morocco.\\
% ...
$^{55}$II. Physikalisches Institut, Georg-August-Universit\"{a}t G\"ottingen, G\"ottingen; Germany.\\
% ...
$^{101}$CPPM, Aix-Marseille Universit\'e, CNRS/IN2P3, Marseille; France.\\
$^{102}$Department of Physics, University of Massachusetts, Amherst MA; United States of America.\\
% ...
$^{116}$Department of Physics, New York University, New York NY; United States of America.\\
% ...
$^{119}$Homer L. Dodge Department of Physics and Astronomy, University of Oklahoma, Norman OK; United States of America.\\
% ...
$^{136}$$^{(a)}$Departamento de F\'isica, Pontificia Universidad Cat\'olica de Chile, Santiago;$^{(b)}$Millennium Institute for Subatomic physics at high energy frontier (SAPHIR), Santiago;$^{(c)}$Instituto de Investigaci\'on Multidisciplinario en Ciencia y Tecnolog\'ia, y Departamento de F\'isica, Universidad de La Serena;$^{(d)}$Universidad Andres Bello, Department of Physics, Santiago;$^{(e)}$Instituto de Alta Investigaci\'on, Universidad de Tarapac\'a, Arica;$^{(f)}$Departamento de F\'isica, Universidad T\'ecnica Federico Santa Mar\'ia, Valpara\'iso; Chile.\\
% ...
$^{149}$Department of Physics, Technion, Israel Institute of Technology, Haifa; Israel.\\
$^{150}$Raymond and Beverly Sackler School of Physics and Astronomy, Tel Aviv University, Tel Aviv; Israel.\\
% ...
$^{a}$ Also Affiliated with an institute covered by a cooperation agreement with CERN.\\
$^{b}$ Also at Borough of Manhattan Community College, City University of New York, New York NY; United States of America.\\
$^{c}$ Also at Bruno Kessler Foundation, Trento; Italy.\\
$^{d}$ Also at Center for High Energy Physics, Peking University; China.\\
$^{e}$ Also at Centro Studi e Ricerche Enrico Fermi; Italy.\\
$^{f}$ Also at CERN, Geneva; Switzerland.\\
% ...

\end{flushleft}


\end{document}